\documentclass[twocolumn,a4paper,preprintnumbers,amsmath,amssymb,nofootinbib,floatfix]{revtex4}
\usepackage{graphicx}% Include figure files 
\usepackage{dcolumn}% Align table columns on decimal point 
\usepackage{bm}% bold math 
\usepackage{epsfig} 
\usepackage{amsfonts}
\usepackage{amsmath}
\usepackage{amssymb}
\usepackage[usenames]{color}
\usepackage[colorlinks=true]{hyperref}
\hypersetup{final=true}
\usepackage{float}
\usepackage{caption}

\begin{document}

\title{J-PAS: forecasts on  dark energy and modified gravity theories}
\author{Miguel Aparicio Resco$^{1}$} \email{migueapa@ucm.es}
\author{Antonio L. Maroto$^{1}$}
\author{Jailson S. Alcaniz$^{2,3}$}
\author{L. Raul Abramo${^4}$}
\author{ C. Hern\'andez-Monteagudo$^{5}$}
\author{N. Ben\'itez$^{6}$}
\author{S. Carneiro$^{7}$}
\author{A. J. Cenarro$^{5}$}
\author{D. Crist\'obal-Hornillos$^{5}$}
\author{R. A. Dupke$^{2,8}$}
\author{A. Ederoclite$^{9}$}
\author{C. L\'opez-Sanjuan$^{5}$}
\author{A. Mar\'in-Franch$^{5}$}
\author{M. Moles$^{5}$}
\author{C. M. Oliveira$^{9}$}
\author{L. Sodr\'e Jr$^{9}$}
\author{K. Taylor$^{10}$}
\author{J. Varela$^{5}$}
\author{H. V\'azquez Rami\'o$^{5}$}

\affiliation{$^{1}$Departamento de F\'{\i}sica Te\'orica and Instituto de F\'isica de Part\'iculas y del Cosmos (IPARCOS), Universidad Complutense de Madrid, 28040 Madrid, Spain}
\affiliation{$^{2}$Observat\'orio Nacional, 20921-400, Rio de Janeiro, RJ, Brasil}
\affiliation{$^{3}$Departamento de F\'{\i}sica, Universidade Federal do Rio Grande do Norte, 59072-970, Natal, RN, Brasil}
\affiliation{$^{4}$Instituto de F\'{\i}sica, Universidade de S\~ao Paulo, R. do Mat\~ao 1371, 05508-090, S\~ao Paulo, SP, Brasil}
\affiliation{$^{5}$Centro de Estudios de F\'{\i}sica del Cosmos de Arag\'on (CEFCA), Plaza de San Juan, 1, E-44001, Teruel, Spain}
\affiliation{$^{6}$Instituto de Astrof\'isica de Andaluc\'ia (CSIC), Glorieta de la Astronom\'ia s/n, Granada, 18008, Spain}
\affiliation{$^{7}$Instituto de F\'isica, Universidade Federal da Bahia, 40210-340, Salvador, BA, Brasil}
\affiliation{$^{8}$Dept. of Astronomy, University of Michigan, 1085 S. University, Ann Arbor}
\affiliation{$^{9}$Instituto de Astronomia, Geof\'isica e Ci\^encias Atmosf\'ericas, Universidade de S\~ao Paulo, R. do Mat\~ao 1226, S\~ao Paulo, SP 05508-090, Brazil}
\affiliation{$^{10}$Instruments4, 4121 Pembury Place, La Ca\~{n}ada-Flintridge, Ca 91011, USA}

\date{\today}

\begin{abstract}
The next generation of galaxy surveys will allow us to test one of the most fundamental assumptions of the standard cosmology, i.e., that gravity is governed by the  general theory of relativity (GR). In this paper we investigate the ability of the Javalambre Physics of the Accelerating Universe Astrophysical Survey (J-PAS) to constrain GR and its extensions. Based on the J-PAS information on clustering and gravitational lensing, we perform a Fisher matrix forecast  on the effective Newton constant, $\mu$, and the gravitational slip parameter, $\eta$, whose deviations from unity would indicate a breakdown of GR. Similar analysis is also performed for the DESI and Euclid surveys and compared to J-PAS with two configurations providing different areas, namely an initial expectation with 4000 $\mathrm{deg}^2$ and the future best case scenario with 8500 $\mathrm{deg}^2$.  We show that J-PAS will be able to measure the parameters $\mu$ and $\eta$ at a sensitivity of $2\% - 7\%$, and will provide the best constraints in the interval $z = 0.3 - 0.6$, thanks to the large number of ELGs detectable in that redshift range. We also discuss the constraining power of J-PAS for dark energy models with a time-dependent equation-of-state parameter of the type $w(a)=w_0+w_a(1-a)$, obtaining  $\Delta w_0=0.058$ and $\Delta w_a=0.24$ for the absolute errors of the dark energy parameters.  

%Constraints on the evolution of the growth function, $f$, and the density-weighted growth rate, $f\sigma_8$, are also derived.

\end{abstract}
\maketitle

%--------------------------------------------------------------------------------------------------------------------------

\section{Introduction}\label{Sec:intro}

The success of the general theory of relativity (GR) is unquestionable. For about a hundred years now, GR has remained unchanged and capable of explaining observations and experiments in a number of regimes, such as the dynamics of the Solar System, gravitational wave emission, the energetics of supermassive black holes and quasars (see e.g. \cite{Will2014} for the status of experimental tests of GR). When extrapolated to cosmological scales, Einstein's theory has also provided a very good description of the evolution of the Universe, which is obtained at the cost of postulating the existence of both dark matter as well as a dark energy component, i.e., an additional field with fine-tuned properties responsible for the current cosmic acceleration~\cite{Sahni:1999gb,Padmanabhan:2002ji,Peebles:2002gy,Copeland:2006wr}. 

Given the unnatural properties of dark energy~\cite{Weinberg:1988cp}, a promising alternative to the standard scenario (GR plus dark energy) is based on infra-red modifications to GR, leading to a weakening of gravity on cosmological scales and thus to  late-time acceleration. In the past few decades, a number of modified or extended theories of gravity  (MG) have been proposed~\cite{Dvali:2000hr,Sahni:2002dx,Capozziello:2002rd,Carroll:2003wy,Santos:2007bs} (see also \cite{Sotiriou:2008rp,Capozziello:2011et,Clifton:2011jh,Ferreira:2019xrr} for recent reviews). In general, these ideas explore as much as they can the loopholes of Lovelock's theorem, while preserving GR on astrophysical scales. Recently, the number of allowed MG theories was significantly restricted~\cite{Baker:2017hug,Creminelli:2017sry,PhysRevLett.119.251304}, given the tight bound on the speed of propagation of gravitational waves, $|c_{\rm{gw}}/c - 1|\lesssim 10^{-15}$, obtained from the 
%recently detected 
binary neutron star merger GW170817~\cite{PhysRevLett.119.161101}. In the near future, other constraints are also expected from black hole imaging, as recently reported by the Event Horizon Telescope\footnote{\url{https://eventhorizontelescope.org}}.

Cosmological observations are also able constrain MG theories at the largest scales, as has been shown by e.g. the Planck experiment \citep{Aghanim:2018eyx}. In this context, the large scale structure surveys that will become available in the coming years will play the major role~\cite{Ferreira:2019xrr}. Those surveys can be categorized in two main types: {\it (i)} spectroscopic surveys, obtaining high-quality spectra (and corresponding high-quality redshift measurements thereof), typically targeting a pre-selected subsample of extragalactic objects (e.g., BOSS \citep{BOSS}, eBOSS \citep{eBOSS}, DESI \citep{DESI,Aghamousa:2016zmz}, Euclid \cite{Laureijs:2011gra,Amendola:2016saw} etc.), and {\it (ii)} photometric surveys, probing the sky at deeper magnitudes in a reduced number of filters, providing significantly larger catalogues of sources, but at the expense of a poorer spectral characterization (e.g. DES \cite{Abbott:2005bi}, LSST 
\cite{Abell:2009aa}, etc). 

An intermediate regime is represented by the so-called spectro-photometric surveys (COMBO-17 \citep{Wolf03}, ALHAMBRA \citep{ALHAMBRAMoles08}, COSMOS \citep{Ilbert09}, MUSYC \citep{Cardamone10}, CLASH \citep{Postman12}, SHARDS \citep{PabloPG13}, PAU \citep{pauS}, J-PLUS \citep{JPLUSCenarro}, J-PAS \citep{Benitez:2014ibt} and SPHEREx \citep{SPHEREX}), that combine deep imaging with multi-color information obtained through combination of broad, medium and narrow band filters. 
 In this way, a low-resolution spectrum (also known as ``pseudospectrum") is obtained for every pixel in the survey's footprint, and in particular for each and all sources present in the joint catalogue extracted from the combination of all bands. This allows providing high-quality photometric redshift estimations for a much larger number of objects compared with spectroscopic surveys, on top of 2D information for those sources that are spatially resolved.

This paper discusses the expected cosmological implications of  J-PAS \cite{Benitez:2014ibt} on dark energy and modified gravity theories.  As is well known, the main body of observations currently available comes from distance measurements which map the expansion history of the Universe at the background level. However, these measurements alone are not enough to discriminate between  a dark energy fluid and modifications to GR, as different models can predict the same expansion history~\cite{Kunz:2012aw}.  Additional observational information is thus required in order to break the model degeneracy and, in particular, the growth of structures and gravitational lensing, which is directly sensitive to the growth of dark matter perturbations -- in contrast with measurements based on galaxies, neutral hydrogren or any other baryonic tracer -- are among the most promising avenues in this respect. 

Here, we consider the J-PAS information on clustering and gravitational lensing and perform a Fisher matrix forecast on the effective Newton constant, $\mu$, and the gravitational slip parameter, $\eta$ (defined in Sec. \ref{Sec:MGmodels}), assuming two configurations of area for J-PAS, i.e.,  4000 $\mathrm{deg}^2$ and 8500 $\mathrm{deg}^2$. For completeness we also discuss the constraining power of J-PAS for dark energy models with a time-dependent equation-of-state parameter $w(a)$, and compare all J-PAS forecasts with those expected by the DESI~\citep{DESI,Aghamousa:2016zmz} and Euclid surveys~\cite{Laureijs:2011gra,Amendola:2016saw}. In this sense, this work updates some of the results contained in \cite{Benitez:2014ibt} and also makes new forecasts, including several new scenarios. Further analysis on interactions in the dark sector can be found in \cite{Costa:2019uvk}.

\section{The J-PAS Survey}

The Javalambre Physics of the Accelerating Universe Astrophysical Survey ({J-PAS}) \cite{Benitez:2014ibt} is a spectro-photometric survey to be conducted at the {Observatorio Astrof\'\i sico de Javalambre} (hereafter OAJ), a site on top of {\it Pico del Buitre}, a summit about $\sim 2,000$\,m high above sea level at the Sierra of Javalambre, in the Eastern region of the Iberian peninsula. The Javalambre Survey Telescope (JST/T250), a 2.5\,m diameter, altazimuthal telescope, will be on charge of J-PAS. JST will be equipped with the Javalambre Panoramic Camera (JPCam), a 14-CCD mosaic camera using a new large format e2v 9.2\,k-by-9.2\,k 10\,$\mu$m pixel detectors, and will incorporate a 54 narrow- and 4 broad-band filter set covering the optical range \citep{JPCam}. The Field of View covered by JPCam is close to 5\,sq.\,deg., and thus the JST/JPCam system constitutes a system specifically defined to optimally conduct spectro-photometric surveys. J-PAS is not the first survey being carried out at the OAJ, since the Javalambre Local Universe Photometric Survey (J-PLUS), conducted by the Javalambre Auxiliary Survey Telescope (JAST/T80), has already covered about 1,600\,sq.\,deg. with 12 broad and narrow band filters (some of them in common to J-PAS). We refer the reader to Benitez et al. \cite{Benitez:2014ibt} and Cenarro et al. \cite{Cenarro:2018uoy} for more details on J-PAS and J-PLUS, respectively.

\section{Dark Energy and Modified Gravity Parameterizations}\label{Sec:MGmodels}

In recent years many different  models of dark energy or MG have been proposed as alternatives to the standard $\Lambda$ -- Cold Dark Matter ($\Lambda$CDM) cosmology. The possibility of confronting such alternatives with observations in a largely model-independent way has motivated the development of theoretical frameworks in which general modifications can be captured in a few effective parameters which can be 
directly tested by observations \cite{Clifton,Silvestri:2013ne}.

In this section we introduce the phenomenological parameterizations of dark energy and MG that will be considered throughout  the paper.

\subsection{Dark Energy}

In the context of GR, dark energy is  understood as a smooth (non-clustering)  energy component  with a sufficient negative pressure, $p$, to violate the strong energy condition ($\rho + 3p \geq 0$, where $\rho$ is the energy density) and accelerate the Universe.  Many different models of dark energy have 
been proposed in recent years (see e.g. \cite{Peebles:2002gy, Copeland:2006wr, Barboza:2008rh} and references therein), based on fluid descriptions with different equations of state or the inclusion of an additional scalar field, as in the quintessence models. 

Rather than focusing on particular models, 
we will consider a phenomenological description of 
dark energy as a perfect fluid with an equation
of state given by the parameterization~\cite{Chevallier:2000qy,Linder:2002et} 
\begin{eqnarray}
w(a) = w_0 + {w_a}(1 - a)\;,
\end{eqnarray}
which reduces to the standard $\Lambda$CDM model for values of $w_0 = -1$ and $w_a = 0$. Note also that this effective modification with respect
to the standard cosmology mainly affects the background evolution.  Notice that the dark energy component could acquire cosmological perturbations  
which are already taken into account in the CAMB code \cite{Lewis:1999bs}.

\subsection{Modified Gravity}

We will consider for simplicity the case of MG theories that include additional scalar degrees of freedom. Extensions of the model-independent approach for modified theories including additional vector fields can be found in \cite{AparicioMarotoVec}. 

Let us then consider the scalar-perturbed flat Friedmann-Lema\^itre-Robertson-Walker (FLRW) metric, written in the longitudinal gauge \cite{amendola2010dark,Tsujikawa:2012hv}:
\begin{eqnarray}
ds^2=-(1+2\Psi)dt^2+a^2(t)(1+2\Phi)d{\bf x}^2\;.
\end{eqnarray}
%where $H(t)=\dot a(t)/a(t)$ and a dot stands for the derivative with respect to $t$. On the other hand, the perturbed  extra scalar degree of freedom reads  $\phi=\phi_0(t)+\delta\phi(t,{\bf x})$. 
The modified Einstein equation to  first order in perturbations can be written as
\begin{eqnarray}
\delta \bar{G}^{\mu}_{\,\,\,\nu}=8 \pi G \, \delta T^{\mu}_{\,\,\,\nu}\;,
\end{eqnarray}
where the perturbed modified Einstein tensor $\delta \bar{G}^{\mu}_{\,\,\,\nu}$  can in 
principle depend on both the metric potentials
$\Phi$ and $\Psi$, and the perturbed scalar  field $\delta\phi$. On the other hand, at late times the only relevant energy component is  non-relativistic matter so that,
\begin{subequations}
\begin{equation}\label{s1.3}
\delta T^{0}_{\,\,\,0}=-\rho_m \, \delta_m,
\end{equation}
\begin{equation}\label{s1.4}
\delta T^{0}_{\,\,\,i}=-\rho_m \, v_{i},
\end{equation}
\begin{equation}\label{s1.5}
\delta T^{i}_{\,\,\,j}=0,
\end{equation}
\end{subequations}
where $v_{i}$ is the three-velocity of matter, $\rho_m$ is the total matter density and  $\delta_m=\delta \rho_m/\rho_m$ is the corresponding  matter  density contrast, which is related to the galaxy density contrast $\delta_g$ via the bias factor $b$, as $\delta_g=b\,\delta_m$.

Using the Bianchi identities in the modified Einstein tensor, we find that in the sub-Hubble regime ($k \gg aH$, $H=\dot a(t)/a(t)$ is the Hubble parameter) there are only two independent Einstein equations, which together with the scalar field equation of motion lead to the following set of equations to first order in perturbations:
\begin{subequations}
\begin{eqnarray}\label{s1.7}
a_{1 1} \, \Psi + a_{1 2} \, \Phi + a_{1 3} \, \delta \phi = -4 \pi G a^2\, \rho_m \, \delta_m,
\end{eqnarray}
\begin{eqnarray}\label{s1.8}
a_{2 1} \, \Psi + a_{2 2} \, \Phi + a_{2 3} \, \delta \phi = 0,
\end{eqnarray}
\begin{eqnarray}\label{s1.9}
a_{3 1} \, \Psi + a_{3 2} \, \Phi + a_{3 3} \, \delta \phi = 0,
\end{eqnarray}
\end{subequations}
where $a_{i j}$ are general differential operators, although for simplicity we will restrict ourselves  to the case of second order operators. In  the quasi-static approximation,  in which  time derivatives can be neglected with respect to the spatial ones,  equations (\ref{s1.7}) - (\ref{s1.9}) are in Fourier space  just algebraic equations
for $(\Phi, \,\Psi, \,\delta \phi)$ in terms of $\delta_m$. 
Notice that the quasi-static approximation is a good one for 
 models with large speed of sound of dark energy perturbations and 
 can be safely employed for current galaxy surveys. For future
 large surveys it could be inappropriate on scales close to the Hubble horizon. Also as shown in \cite{Sawicki:2015zya} it should never be used for the integrated Sachs-Wolfe effect analysis.  

By eliminating the 
scalar degree of freedom from (\ref{s1.9}) and 
substituting it into (\ref{s1.7}) and (\ref{s1.8}), we obtain 
two effective equations for the metric perturbations, which can be written as
\begin{eqnarray}\label{s1.12}
k^2 \, \Phi = 4\pi Ga^2 \,\mu \, \eta \, \rho_m \delta_m,
\end{eqnarray}
\begin{eqnarray}\label{s1.13}
k^2 \, \Psi = -4\pi G a^2 \mu\, \rho_m \delta_m.
\end{eqnarray}
Note that on the sub-Hubble scales, 
$\delta_m$ agrees with the density perturbation $\Delta$ used in \cite{Silvestri:2013ne} since $\Delta=\delta_m+\frac{3 aH v}{k}$. Therefore,  in the quasi-static approximation, a general modification of Einstein's equations can be written in terms of two arbitrary functions of time and scale $\mu(a,k)$ and $\eta(a,k)$ \cite{Pogosian:2010tj, Silvestri:2013ne}. These parameters can be understood  as an effective Newton constant, $G_{\text{eff}}(a,k)$, given by
\begin{eqnarray}\label{s1.16}
\mu(a,k)=\frac{G_{\text{eff}}}{G},
\end{eqnarray}
and the gravitational slip parameter  
\begin{eqnarray}\label{s1.17}
\eta(a,k)=-\frac{\Phi}{\Psi}, 
\end{eqnarray}
which modifies the equation for the lensing potential, that depends upon the combination $(\Psi-\Phi)/2$. Thus, deviations from $\mu = \eta = 1$ indicate a breakdown of standard GR.

The modified equations can be rewritten as 
\begin{eqnarray}
k^2\Psi\simeq -4\pi G_{\text{eff}}\,a^2\rho_m\delta_m\;,
\end{eqnarray}
and
\begin{eqnarray}
\frac{\Psi-\Phi}{2}\simeq -\frac{3G_{\text{eff}}}{2G}\frac{1+\eta}{2}\left(\frac{aH}{k}\right)^2\Omega_m(a)\delta_m \label{lensing}\;.
\end{eqnarray}
where $\Omega_m (a) = \Omega_m \, a^{-3} \, E(a)^{-2}$ is the matter density parameter and $E(a) = H(a)/H_0$, with the Hubble constant written as $H_0 = 100h\, \rm{km \, s^{-1} \, Mpc^{-1}}$. 

Using the standard conservation equation,  $T^{\mu\nu}_{\; ;\nu}=0$, we obtain the continuity and Euler equations, which in the sub-Hubble regime and for 
non-relativistic matter, reduce to
\begin{eqnarray}
a\dot\delta_m&=&-\theta\;, \label{cont}\\
a\dot\theta&=& -aH\theta+k^2\Psi\;, \label{Euler}
\end{eqnarray}
where $\theta=i ({\bf k}\cdot {\bf v})$. 

Taking the time derivative of (\ref{cont}) and using (\ref{Euler}), we obtain the modified growth equation which reads
 \begin{eqnarray}
\delta''_m+\left(2+\frac{H'}{H}\right)\delta'_m-\frac{3}{2}\mu(a,k)\Omega_m(a)\delta_m\simeq 0 \label{growth}\;,
\end{eqnarray}
where the prime denotes derivative with respect to $\ln a$.  In general, it can be shown that for any local and generally covariant four-dimensional theory of gravity, the effective parameters $\mu(a,k)$ and $\eta(a,k)$ reduce to rational functions of $k$. % which are even in theories with purely scalar extra degrees of freedom. 
If we also assume that higher than second derivatives do not appear in the equations of motion, then they can be completely described by five functions of time only as follows \cite{Silvestri:2013ne}:
\begin{eqnarray}
\mu(k,a)=\frac{1 + p_3(a)k^2}{p_4(a) + p_5(a)k^2}\;,
\end{eqnarray}
and 
\begin{eqnarray}
\eta(k,a)=\frac{p_1(a) + p_2(a)k^2}{1 + p_3(a)k^2}\;.
\end{eqnarray}
Notice that in general the $p_i(a)$ parameters remain essentially 
unconstrained by the GW170817~\cite{PhysRevLett.119.161101}
event observation, although in some particular 
frameworks such as those of Hordenski theories, 
the condition $c_t=1$ imposes certain constraints
on the parameters \cite{Kase:2018aps}. 
For simplicity, in our analysis we will 
limit ourselves to two particular classes of effective parameters, namely scale-independent parameterizations with $\mu=\mu(a)$ and $\eta=\eta(a)$ and time-independent parameterizations, i.e., $\mu=\mu(k)$ and $\eta=\eta(k)$. In the scale-independent case, two particularly relevant examples will be analized. On one hand,  the constant in time case and, on the other, the parameterization proposed in \cite{Simpson:2012ra}, which is usually employed in  the literature  \cite{Ade:2015rim},
\begin{eqnarray}\label{31}
\mu (a) = 1+(\mu_0-1) \, \frac{1-\Omega_m (a)}{1-\Omega_m},
\end{eqnarray}
\begin{eqnarray}\label{32}
\eta (a) = 1+(\eta_0-1) \, \frac{1-\Omega_m (a)}{1-\Omega_m}\;.
\end{eqnarray}
%
%where $\Omega_m (a) = \Omega_m \, (1+z)^3 \, E(z)^{-2}$. 
This parameterization ensures that at high redshift the standard GR values
are recovered.

%A different parametrization $(\mu,\Sigma)$ has been also used with:
%\begin{eqnarray}
%\Sigma=\frac{\mu}{2}(1+\eta)\;.
%\end{eqnarray}
%We now provide a list of the MG models that we are going to address. From the point of view of perturbations, their analysis can be tackled with a model-independent approach via the two independent functions $\mu(a,k)$ and  $\eta(a,k)$ introduced above. Observables built out of the matter power spectrum in redshift space and the lensing convergence power spectrum will allow to compute the Fisher matrices for $\mu(a,k)$ and  $\eta(a,k)$ in different redshift bins and different $k$ bins.

\section{Fisher Matrices for Galaxy and Lensing Power Spectra}\label{Sec:Fishermatrix}
The Fisher matrix formalism provides a simple way to estimate the precision 
with which certain cosmological parameters could be measured from a set of 
observables once the survey specifications and the fiducial cosmology are fixed.
Thus, given a set of parameters $\{p_\alpha\}$, the Fisher matrix $\textbf{F}^{p}$ 
is just the inverse of the covariance matrix in the parameters space. It provides the 
marginalized error for the $p_\alpha$ parameter as $\sqrt{F^{-1}_{\alpha\alpha}}$. The corresponding $1\sigma$ region is just an ellipsoid in the parameter space
since  the probability distribution function (PDF) are asummed to be Gaussian in the Fisher formalism.
If we are interested in obtaining errors for a different set of parameters $\{q_\alpha\}$, the Fisher matrix of the new parameters simply reads, 
\begin{equation}\label{25a}
\textbf{F}^{q}=\textbf{P}^{t} \, \textbf{F}^{p} \, \textbf{P},
\end{equation}
where $\textbf{P}=\textbf{Q}^{-1}$ and $Q_{\alpha\beta}=\partial{q_{\alpha}}/\partial{p_{\beta}}$, evaluated on the fiducial model. 

In the following, we provide general expressions for the Fisher matrices for the galaxy power spectrum in redshift space and for the lensing convergence power spectrum, both in different redshift and $k$ (or $\ell$) bins.
We will apply them  separately to J-PAS \cite{Benitez:2014ibt}, DESI \cite{Aghamousa:2016zmz} and Euclid \cite{Laureijs:2011gra} galaxy surveys and for 
J-PAS and Euclid lensing surveys.

\subsection{Fisher Matrix for Galaxy Clustering}

Following  \cite{Amendola:2012ky, Amendola:2013qna}, let us introduce the following dimensionless parameters  $A$ and $R$,
\begin{equation}\label{2.1}
A=D \, b \, \sigma_{8},
\end{equation}
\begin{equation}\label{2.2}
R=D \, f \, \sigma_{8},
\end{equation}
where $D(z)=\delta_m(z)/\delta_m(0)$ is the growth factor, $b$ is the bias and $f$ is the growth function defined by
\begin{equation}\label{2.3}
D(z)=\exp\left[\int_{0}^{N(z)} f(N') \, dN'\right],
\end{equation}
being $N(z)=-\log(1+z)$.
The  $\sigma_{8}$ constant corresponds to $\sigma_8=\sigma(0.8 \, \textrm{Mpc}/\textrm{h})$ where,
\begin{eqnarray}\label{2.4}
\sigma^{2}(z,R)= D^{2}(z) \int{\frac{k'^{2}\, dk'}{2\pi^{2}} P(k') |\hat{W}(R,k')|^{2}} ,
\end{eqnarray}
being $P(k)$ the matter power spectrum.  We use a top-hat filter $\hat{W}(R,k)$, defined by
\begin{equation}\label{2.5}
\hat{W}(R,k)=\frac{3}{k^{3}R^{3}} \, [\sin(kR)-kR\cos(kR)].
\end{equation}

Then, the galaxy power spectrum in redshift space is \cite{Seo:2003pu},
\begin{equation}\label{2.7}
P(k_{r},\hat{\mu}_{r},z)=\frac{D_{A \,r}^{2} \, E}{D_A^{2} \, E_{r}} \, (A+R \, \hat{\mu}^{2})^{2} \, \hat{P}(k) \, e^{-k_{r}^{2} \, \hat{\mu}_{r}^{2} \, \sigma_{r}^{2}},
\end{equation}
where  sub-index $r$ denotes that the corresponding quantity is evaluated on the fiducial model, $\hat{P}(k) \equiv P(k)/\sigma_8^2$, $\sigma_{r}=(\delta z \, (1+z))/H(z)$ with $\delta z(1+z)$ the photometric redshift error, and $D_A$ is the angular distance which, in a flat Universe, reads $D_A=(1+z)^{-1} \, \chi (z)$, with 
\begin{equation}\label{2.8}
\chi(z)=H_{0}^{-1} \, \int_{0}^{z} \frac{dz'}{E(z')}.
\end{equation}
The dependences $k=k(k_{r})$, $\hat{\mu}=\hat{\mu}(\hat{\mu}_{r})$ and the factor $\frac{D_{Ar}^2 \, E}{D_A^2 \, E_{r}}$ are due to the Alcock-Paczynski effect \cite{Alcock:1979mp} (see e.g. \cite{amendola2010dark})% pages $393 \, - \, 394$),
\begin{equation}\label{2.9}
k=Q \, k_{r},
\end{equation}
\begin{equation}\label{2.10}
\hat{\mu}=\frac{E \, \hat{\mu}_{r}}{E_{r} \, Q},
\end{equation}
\begin{equation}\label{2.11}
Q=\frac{\sqrt{E^{2} \, \chi^{2} \, \hat{\mu}^{2}_{r}-E_{r}^{2} \, \chi_{r}^{2} \, (\hat{\mu}^{2}_{r}-1)}}{E_{r} \, \chi}.
\end{equation}

If we consider different galaxies as dark matter tracers with bias $b_i$, the galaxy power spectrum is \cite{White:2008jy, McDonald:2008sh},
\begin{eqnarray}\label{2.12}
P_{i j}(k_{r},\hat{\mu}_{r},z) & = & \frac{D_{A \,r}^{2} \, E}{D_A^{2} \, E_{r}} \, (A_i+R \, \hat{\mu}^{2})   \\ & & \times  (A_j+R \, \hat{\mu}^{2}) \hat{P}(k) \, e^{-k_{r}^{2} \, \hat{\mu}_{r}^{2} \, \sigma_{r}^{2}}, \nonumber
\end{eqnarray}
where $A_i = D \, b_i \, \sigma_{8}$. Then, considering a set of cosmological parameters $\{p_\alpha\}$, the corresponding Fisher matrix for clustering of different tracers and for a given redshift bin centered at $z_a$ is \cite{Abramo:2011ph, Abramo:2015iga},
\begin{eqnarray}\label{2.13}
&&F_{\alpha \beta}^{C} (z_a)  =  \frac{V_a}{8 \pi^{2}} \int_{-1}^{1} d\hat{\mu} \int_{k_{\text{min}}}^{\infty} dk \,\, k^{2} \, \left.\frac{\partial P_{i j}(k,\hat{\mu},z_a)}{\partial p_\alpha}\right|_{r}  \,  \\ & & \times C^{-1}_{j l} \, \left.\frac{\partial P_{l m}(k,\hat{\mu},z_a)}{\partial p_\beta}\right|_{r}  \, C^{-1}_{m i} \,\,\, \mathrm{e}^{-k^2 \, \Sigma^{2}_{\perp}-k^2 \, \hat{\mu}^2 \, (\Sigma^2_{\parallel}-\Sigma^2_{\perp})}\;, \nonumber
\end{eqnarray}
where 
\begin{equation}\label{2.14}
\Sigma_{\perp}(z) = 0.785 \, D(z) \, \Sigma_0,
\end{equation}
\begin{equation}\label{2.15}
\Sigma_{\parallel}(z) = 0.785 \, D(z) \, (1+f(z)) \, \Sigma_0,
\end{equation}
with $\Sigma_0 = 11h^{-1} \mathrm{Mpc}$ for our fiducial value of $\sigma_8=0.82$ in the modified gravity case, and $\Sigma_0 = 6.5 h^{-1} \mathrm{Mpc}$ for the dark energy case due to the reconstruction procedure \cite{Seo:2007ns}. Finally $k_{\text{min}}$ is fixed to 0.007 $h$/Mpc \cite{Amendola:2013qna}.  Thus the exponential cutoff \cite{Seo:2007ns} removes the contribution from non-linear scales across and along the line of sight. The factor 0.785 takes into account the different normalization of $(1+z) \, D(z)$ at high redshifts compared to \cite{Seo:2007ns} \footnote{Note that there is a typo in the normalization factor 0.785 on \cite{Seo:2007ns}. We thank C\'assio Pigozzo for pointing this out.}. The 
data covariance matrix is
\begin{equation}\label{2.16}
C_{i j} = P_{i j} + \frac{\delta_{i j}}{\bar n_i},
\end{equation}
where $\bar n_i = \bar n_i (z_a)$ is the mean galaxy density of tracer $i$ in the $z$ bin $a$.
Finally, $V_{a}$ is the total volume of the $\textrm{a}$-th bin. For a flat $\Lambda$CDM model,  $V_{a}=\frac{4 \pi \, f_{sky}}{3} \, \left(\chi(\bar{z}_{a})^{3}-\chi(\bar{z}_{a-1})^{3}\right)$ where $f_{sky}$ is the sky fraction of the survey and $\bar{z}_{a}$ the upper limit of the $a$-th bin. 
For the particular case in which we have only one tracer we recover from (\ref{2.13}) the standard Fisher matrix of clustering for the power spectrum (\ref{2.7}) at $z_a$ \cite{Seo:2003pu}, 
\begin{eqnarray}\label{2.20}
&&F_{\alpha \beta}^{C} (z_a)  =  \frac{V_a}{8 \pi^{2}} \int_{-1}^{1} d\hat{\mu} \int_{k_{\text{min}}}^{\infty} k^{2} \, V_{eff} \, \left.\frac{\partial \ln(P(k,\hat{\mu},z_a))}{\partial p_\alpha}\right|_{r}  \nonumber \\ & & \times \left.\frac{\partial \ln(P(k,\hat{\mu},z_a))}{\partial p_\beta}\right|_{r}  \mathrm{e}^{-k^2 \, \Sigma^{2}_{\perp}-k^2 \, \hat{\mu}^2 \, (\Sigma^2_{\parallel}-\Sigma^2_{\perp})} \, dk\;.
\end{eqnarray}
where 
%$V_{eff}=V_{eff}(k,\hat{\mu},z_{a})|_{r}$,
$V_a$ is the volume of the redshift slice $z_a$, and the effective volume is given by,
\begin{equation}\label{2.21}
V_{eff}=\left(\frac{\bar n(z_{a}) \, P(k,\hat{\mu},z)}{1+\bar n(z_{a}) \, P(k,\hat{\mu},z)}\right)^{2}.
\end{equation}

Finally, if we are interested in estimating errors in 
different $k$-bins, we  sum the information for all $z$ bins in each $k_q$ bin of width $\Delta k_q$, so that
\begin{eqnarray}\label{2.22}
&&F_{\alpha \beta}^{C} (k_q)  =  \sum_a \frac{V_a}{8 \pi^{2}} \int_{-1}^{1} d\hat{\mu} \int_{\Delta k_{q}} dk \, k^{2} \, \left.\frac{\partial P_{i j}(k,\hat{\mu},z_a)}{\partial p_\alpha}\right|_{r} \nonumber \\ & &  \times C^{-1}_{j l} \left.\frac{\partial P_{l m}(k,\hat{\mu},z_a)}{\partial p_\beta}\right|_{r}  C^{-1}_{m i} \mathrm{e}^{-k^2  \Sigma^{2}_{\perp}-k^2  \hat{\mu}^2 (\Sigma^2_{\parallel}-\Sigma^2_{\perp})}. 
\end{eqnarray}
 \begin{figure*}[ht!]
\begin{center}
 \includegraphics[width=0.45\textwidth]{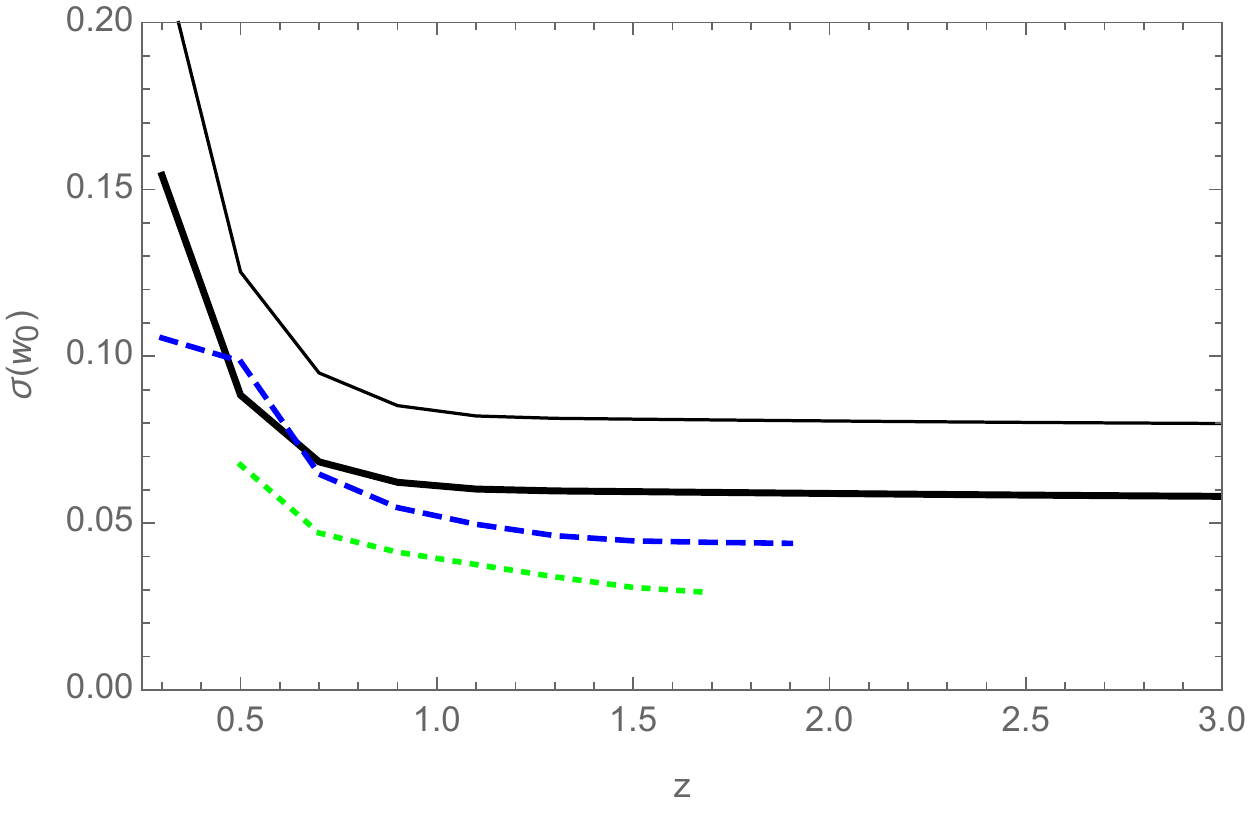} 
\includegraphics[width=0.45\textwidth]{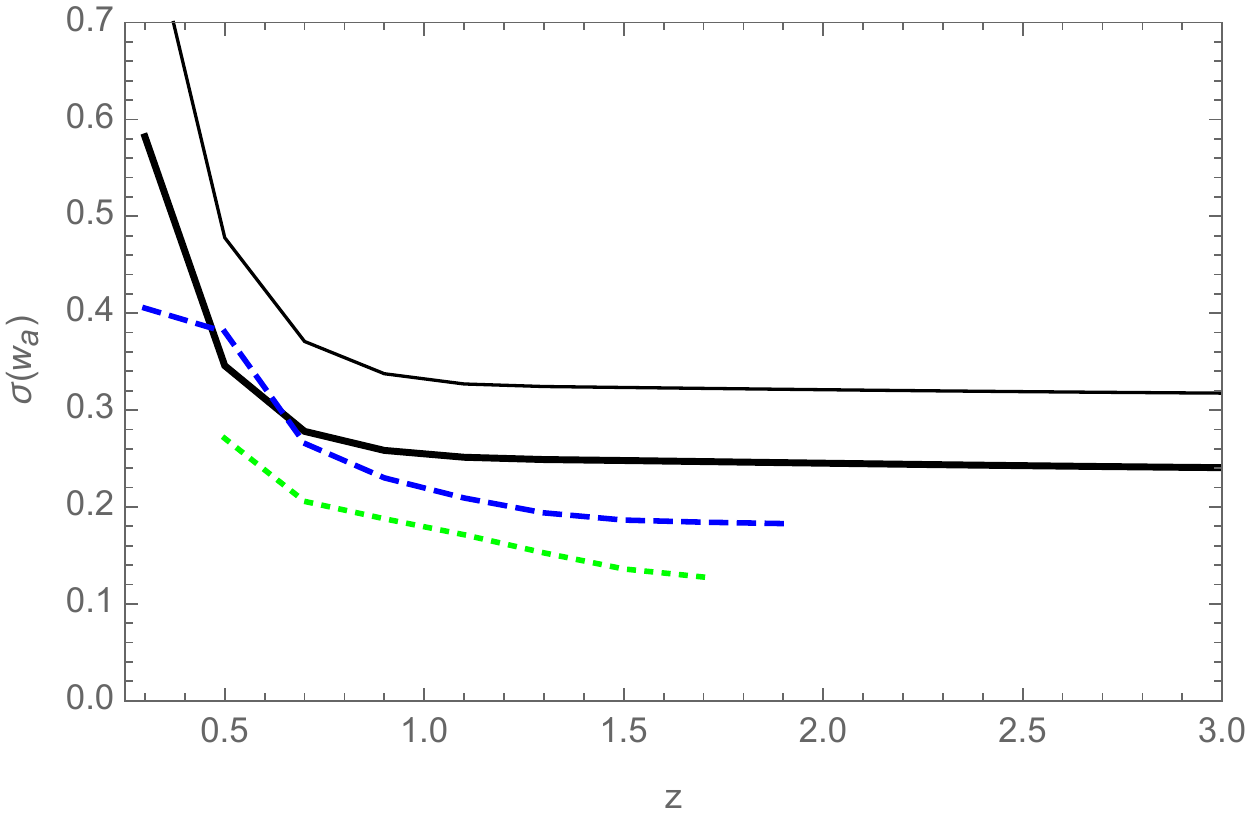} 
 \includegraphics[width=0.48\textwidth]{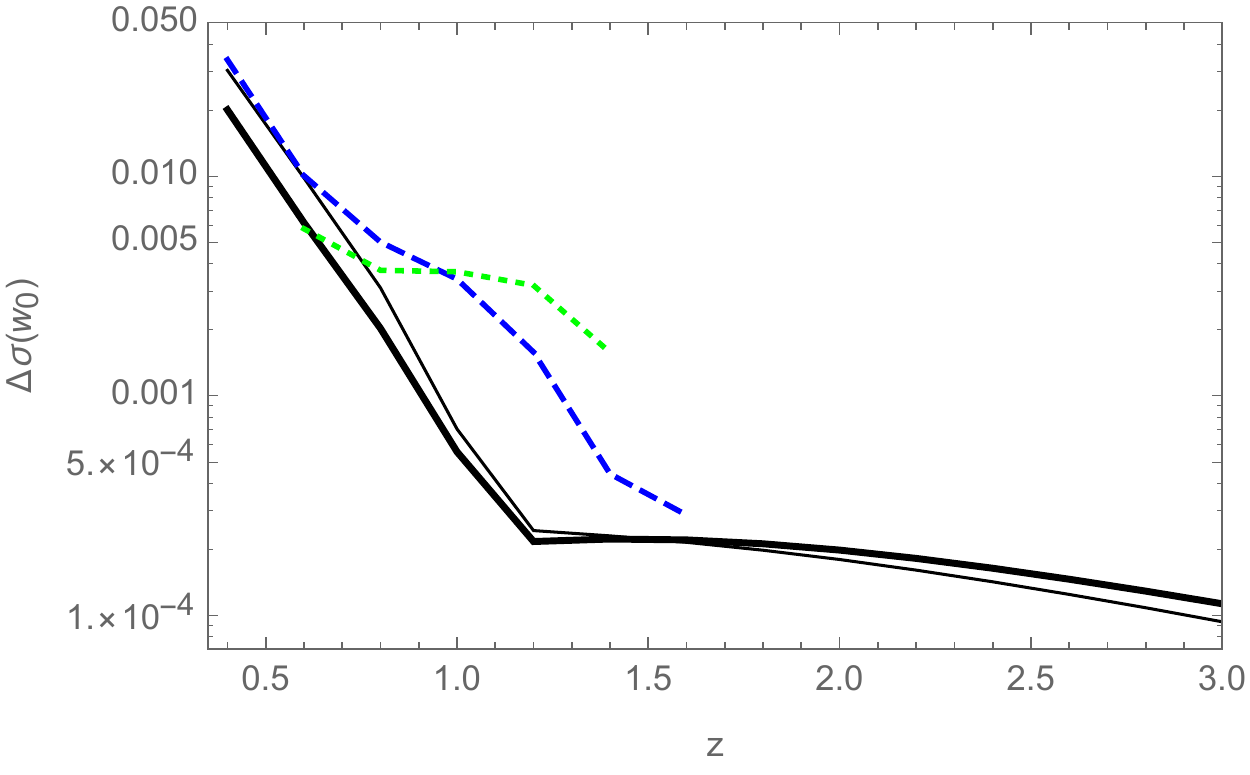} 
\includegraphics[width=0.48\textwidth]{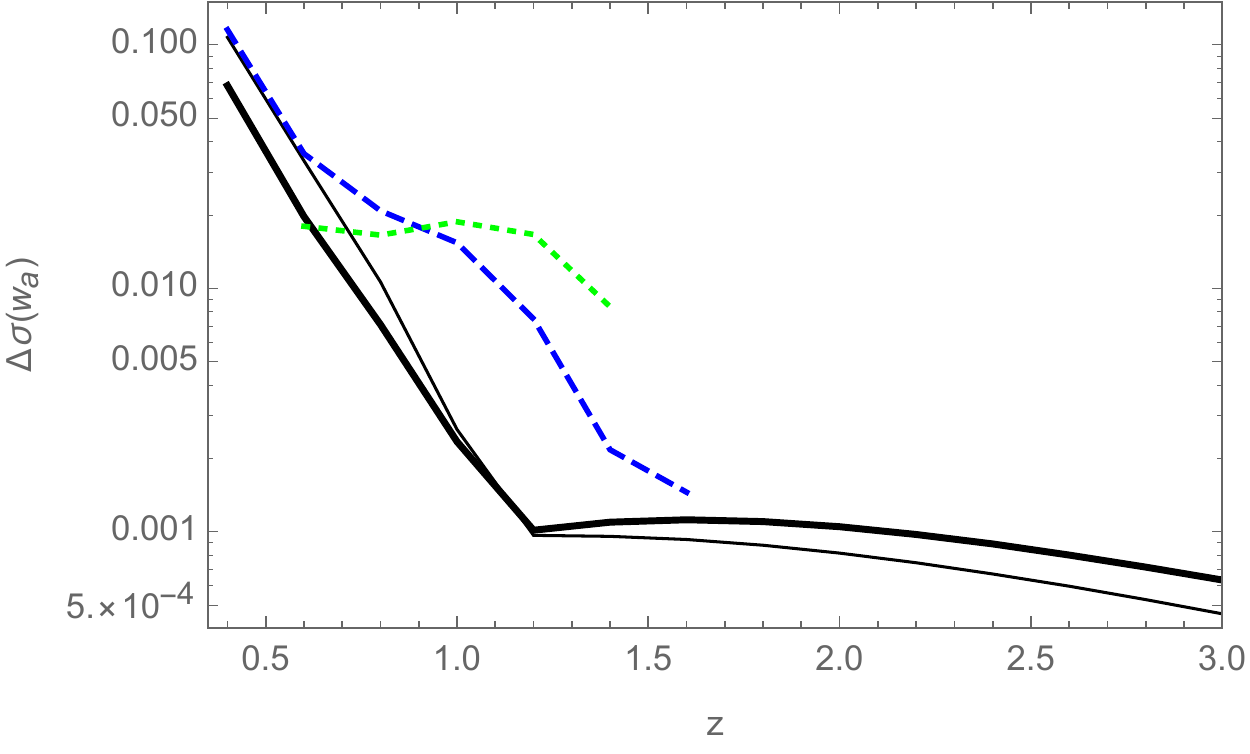}
 \end{center}
\caption{Top panel: constraints on $w_0$ (left) and $w_a$ (right) as we increase the depth of the surveys. Here we consider only the clustering  information. The errors for J-PAS (8500 deg$^2$, black solid lines; 4000 deg$^2$, black thin lines) combine ELGs, LRGs and QSOs; those for DESI (14000 deg$^2$, blue dashed lines) combine the BGS sample, ELGs, LRGs and QSOs; and those for Euclid (15000 deg$^2$, green dotted lines) include only ELGs. Bottom panel: added value of each successive redshift slice (assuming here bins of $\Delta z =0.2$).}
 \label{fig:w0wa} 
\end{figure*}

\begin{figure*}%[h!]
 \begin{center}
 \includegraphics[width=0.6\textwidth]{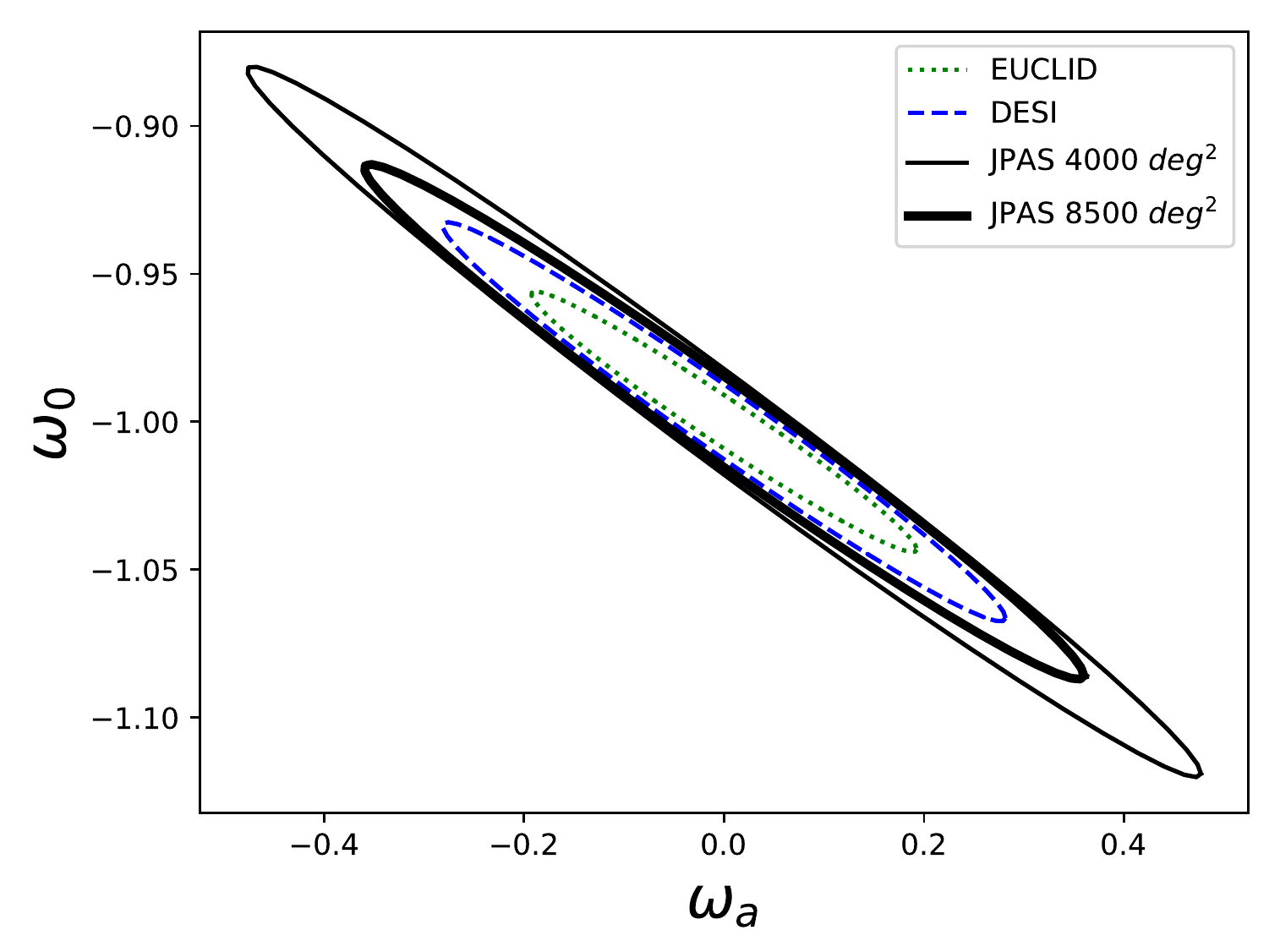}
  \end{center}
  \caption{1$\sigma$ contour error for $w_0$ and $w_a$ for J-PAS (8500 deg$^2$, black solid lines; 4000 deg$^2$, black thin lines) combine ELGs, LRGs and QSOs; those for DESI (14000 deg$^2$, blue dashed lines) combine the BGS sample, ELGs, LRGs and QSOs; and those for Euclid (15000 deg$^2$, green dotted lines) include only ELGs.}
 \label{Figure_2} 
 \end{figure*}
\subsection{Fisher Matrix for Weak Lensing}\label{len}
%%%%%%%%%%%%%%%%%%%%%%%%%%%%%%%%%%%%%%%%%%%%%%%%%%%%%%%%%%%%%%%%%%%%%%%%%%%%%%%%%%%%

The main observable for the weak lensing measurements is the convergence power spectrum. Using the Limber and flat-sky approximations  we obtain~\cite{Lemos:2017arq}
\begin{equation}\label{16a}
P(\ell) = \int_0^{\infty} \, dz \, \frac{H_0^2 \Omega_m^2}{H(z)} \, K^2(z) \, \frac{\mu^2 \, (1+\eta)^2}{4} \, D^2(z) \, P\left( \frac{\ell}{\chi(z)} \right),
\end{equation}
where $K(z)$ is defined as
\begin{equation}\label{16b}
K(z) = \frac{3\,H_{0}}{2} \, (1+z) \, \int_z^{\infty} \, \left( 1-\frac{\chi(z)}{\chi(z')} \right) \, n(z') \, dz',
\end{equation}
being $n(z)$ the source galaxy density function as a function of the redshift. 
For a redshift tomography analysis, we can generalize the convergence power spectrum as \cite{Hu:1999ek},
\begin{equation}\label{17}
P_{ij}(\ell)\simeq H_{0} \sum_{a} \frac{\Delta z_{a}}{E_{a}} K_{i} (z_{a}) K_{j} (z_{a}) L_{a}^{2}  \hat{P}\left( \frac{\ell}{\chi(z_{a})} \right),
\end{equation}
where we have discretized the integral (\ref{16a}) and defined the dimensionless parameter $L$ as \cite{Amendola:2012ky}
\begin{equation}\label{16}
L=\Omega_{m} \, D \, \frac{\mu \, (1+\eta)}{2} \, \sigma_{8},
\end{equation}
where $L_a=L(z_a)$. The function $K_{i}$ is related to the weak lensing window function for the $i$-bin by
\begin{equation}\label{18}
K_{i} (z)=\frac{3\,H_{0}}{2} \, (1+z) \, \int_{z}^{\infty} \left( 1-\frac{\chi(z)}{\chi(z')} \right) \, n_{i} (z') \, dz',
\end{equation}
where $n_{i} (z)$ is the density function for the $i$-bin,
which is obtained as follows: let us first consider the source galaxy density function for the survey \cite{Ma:2005rc},
\begin{equation}\label{19}
n(z) = \frac{3}{2 z_{p}^{3}} \, z^{2} \, e^{-(z/z_{p})^{3/2}},
\end{equation}
where $z_{p}=z_{mean}/\sqrt{2}$, being $z_{mean}$ the survey mean redshift. Then, within the  $i$-bin we have a new distribution function which is defined to be equal to $n(z)$ inside the bin and zero outside. Now, taking into account the  photometric redshift error, $\sigma_{i}=\delta z \, (1+ z_{i})$,  we obtain
\begin{equation}\label{20}
n_{i}(z) \propto \int_{\bar{z}_{i-1}}^{\bar{z}_{i}} z'^{2} e^{-(z'/z_{p})^{3/2}} \, e^{\frac{(z'-z)^{2}}{2 \sigma_{i}^{2}}} dz',
\end{equation}
%
%where the  photometric error can be written as $\sigma_{i}=\delta z \, (1+ z_{i})$ and $\bar{z}_{i}$ is the upper limit of the $i$-bin. 
%Finally we have to normalize $n_{i}(z)$ properly. 
where $\bar{z}_{i}$ is the upper limit of the $i$-bin.
Then, the Fisher matrix for weak lensing is given by \cite{Eisenstein:1998hr},
\begin{equation}\label{21}
F_{\alpha \beta}^{L}=f_{sky}  \sum_{\ell} \Delta \ln \ell  \frac{(2\ell+1)  \ell}{2}  \textup{Tr}\left[ \frac{\partial \textbf{P}}{\partial p_{\alpha}} \textbf{C}^{-1} \frac{\partial \textbf{P}}{\partial p_{\beta}} \textbf{C}^{-1}\right],
\end{equation}
where $\textbf{P}$ and $\textbf{C}$ are the matrix of size $n_{b} \, \times \, n_{b}$ with,
\begin{equation}\label{22}
C_{ij}=P_{ij}+\gamma_{int}^{2} \, \hat{n}_{i}^{-1} \, \delta_{ij},
\end{equation}
$\gamma_{int} = 0.22$ being the intrinsic ellipticity (see for instance \cite{Hilbert:2016ylf}). Notice that we are not considering the effect of possible systematic errors in 
the shear measurements \cite{Huterer:2005ez}. Finally,  
$\hat{n}_{i}$ denotes the number of galaxies per steradian in the $i$-th bin,
\begin{equation}\label{23}
\hat{n}_{i}=n_{\theta} \, \frac{\int_{\bar{z}_{i-1}}^{\bar{z}_{i}} n(z) \, dz}{\int_{0}^{\infty} n(z) \, dz},
\end{equation}
where $n_{\theta}$ is the areal galaxy density. We sum in $\ell$ with $\Delta \ln \ell=0.1$ from $\ell_{\text{min}}=5$ \cite{Amendola:2013qna} to $\ell_{\text{max}}$ with
$\ell_{\text{max}}=\chi(z_{\alpha'}) \, k_{\text{max}}$ where $\alpha' = \mathrm{min}(\alpha,\beta)$ and $k_{\text{max}}(z_{a})$
is defined so that $\sigma(z_{a},\pi/2k_{\text{max}}(z_{a}))=0.35$ using (\ref{2.4}), i.e. we only consider modes in the linear regime. 

Finally, if we are interested in estimating errors in 
different $\ell$-bins, we introduce a window function in the Fisher matrix (\ref{21}) in order to take into account only the information of a bin $\ell_a$ of width $\Delta \ell_a$,
\begin{eqnarray}\label{21b}
F_{\alpha \beta}^{L} (\ell_a)&=&f_{sky} \, \sum_{\ell} \Delta \ell \, \frac{(2\ell+1)}{2} \,\, W_{a}(\ell) \nonumber \\ & & \times \textup{Tr}\left[ \frac{\partial \textbf{P}}{\partial p_{\alpha}} \textbf{C}^{-1} \frac{\partial \textbf{P}}{\partial p_{\beta}} \textbf{C}^{-1}\right],
\end{eqnarray}
where $W_{a}(\ell)$ is defined as
\begin{equation}\label{21c}
W_{a}(\ell)=\theta\left(\ell - \left[ \ell_a - \frac{\Delta \ell_a}{2} \right]\right) \theta\left(\left[ \ell_a + \frac{\Delta \ell_a}{2} \right]-\ell\right),
\end{equation}
being $\theta(x)$ the Heaviside function.

\subsection{Fiducial Model and Surveys Specifications}

The fiducial J-PAS cosmology \cite{Costa:2019uvk} assumed in our analysis is the flat \textrm{$\Lambda$CDM} model with the parameters $\Omega_m=0.31$, $\Omega_{b}=0.049$, $n_{s}=0.96$, $h=0.68$, $H^{-1}_{0}=2997.9 \, \textrm{Mpc/h}$, and $\sigma_{8}=0.82$ which are compatible with Planck 2018 \cite{Aghanim:2018eyx}.
For this cosmology, the $E(z)$ function defined previously is given by
\begin{equation}\label{4}
E(z)=\sqrt{\Omega_{m} \, (1+z)^{3}+(1-\Omega_{m})}\;,
\end{equation}
whereas the growth function can be written as
\begin{equation}\label{3}
f_{\Lambda}(z)=\left(\Omega_{m} \, (1+z)^{3} \frac{1}{E^2(z)} \right)^{\gamma},
\end{equation}
with   the growth index $\gamma = 0.545$ \cite{Linder:2007hg}. For the fiducial cosmology, the linear matter power spectrum $\hat{P} (k)$ takes the form
\begin{equation}\label{15a}
\hat{P} (k) \propto k^{n_{s}} \, T^2(k),
\end{equation}
where the transfer function has been obtained from CAMB \cite{Lewis:1999bs}. Then, we impose the  normalization
\begin{equation}\label{15b}
\int{\frac{k'^{2}\, dk'}{2\pi^{2}} \hat{P}(k') |\hat{W}(8 \, \textrm{Mpc/h},k')|^{2}}=1,
\end{equation}
since we have taken out $\sigma_{8}^{2}$ from the power spectrum and have inserted it in the definitions ($\ref{2.1}$) and ($\ref{2.2}$). In the dark energy case, we will consider derivatives of the transfer function with respect to $w_0$ and $w_a$ parameters when 
calculating the corresponding Fisher matrices. However in the modified gravity case this is no longer as the 
dependence of the transfer functions on the  modified gravity 
parameters is not explicitly known. For the bias, we consider four different types of galaxies: Luminous Red Galaxies (LRGs), Emission Line Galaxies (ELGs), Bright Galaxies (BGS) ans quasars (QSO) \cite{Mostek:2012nc, Ross:2009sn}.
Each type has different fiducial bias given by
\begin{equation}\label{28}
b(z)=\frac{b(0)}{D(z)},
\end{equation}
being $b_{0}=0.84$ for ELGs, $b_{0}=1.7$ for LRGs and $b_{0}=1.34$ for BGS. For Euclid survey we use a fiducial bias for ELGs of the form 
%
%\begin{equation}\label{29}
$b(z)=\sqrt{1+z}$
%\end{equation}
%
\cite{Laureijs:2011gra}, while the bias for quasars is
%
%\begin{equation}\label{29a}
$b(z)=0.53+0.289 \, (1+z)^{2}$.
%\end{equation}

Finally, we summarize the surveys specifications necessary to compute the different Fisher matrices. For clustering we have considered: redshift bins and galaxy densities for each bin which can be found in the left panel of Table \ref{taJPASDESI} for J-PAS, in the center panel of Table \ref{taJPASDESI} for DESI  and in the right panel of Table \ref{taJPASDESI} for Euclid. We consider two configurations of total area for J-PAS, namely  8500 $\mathrm{deg}^2$ and 4000 $\mathrm{deg}^2$ which correspond to  fractions of the sky of $f_{sky}=0.206$  and $f_{sky}=0.097$ respectively. $f_{sky}=0.339$ for DESI with 14000 $\mathrm{deg}^2$ and $f_{sky}=0.364$ for Euclid with 15000 $\mathrm{deg}^2$. The redshift error is $\delta z=0.003$ for galaxies and QSO in J-PAS, $\delta z=0.0005$ for galaxies in DESI  and $\delta z=0.001$
for QSO in DESI and galaxies in Euclid. 

For the weak lensing analysis we have used: redshift bins and the fraction of the sky $f_{sky}$, which are the same as in the clustering analysis; mean redshifts  for the galaxy density which are $z_{mean}=0.5$ for J-PAS and $z_{mean}=0.9$ for Euclid; the angular number density $n_{\theta}$ (in galaxies per square arc minute) which can be found in Table \ref{ta5} for J-PAS with three different photometric errors.  For Euclid, $n_{\theta}=35$ galaxies per square arc minute with $\delta z=0.05$.

%%%%%%%%%%%%%%%%%%%%%%%%%%%%%%%%%%%%%%%%%%%%%%%%%%%%%%%%%%%%%%%%%%%%%%%%%%%%%%%%%%%%
\section{Results}
%%%%%%%%%%%%%%%%%%%%%%%%%%%%%%%%%%%%%%%%%%%%%%%%%%%%%%%%%%%%%%%%%%%%%%%%%%%%%%%%%%%%

\subsection{Galaxy Clustering}

\subsubsection{Dark Energy}

The dark energy equation of state is one of the main drivers of modern galaxy surveys. Low-redshift measurements of the scale of baryonic acoustic oscillations (BAOs) in galaxy clustering constitute a straightforward, nearly systematic-free way of measuring distances using the ``cosmic standard ruler'' provided by the acoustic horizon at the epoch of baryon drag \cite{Seo:2003pu}. These distances are measured both along the line of sight (since $d\chi = c dz/H(z)$) as well as across the line of sight (using the angular-diameter distance, which for an object of size $dL$ subtending an angle $d\theta$ reads $d\theta = dL/D_A$). The different dependencies of $H(z)$ and $D_A(z)$ on cosmological parameters help break degeneracies, improving the constraints.

In order to derive these constraints, the BAOs derived from galaxy clustering must be compared against the high-redshift measurement of the acoustic horizon from observations of the cosmic microwave background \cite{Ade:2015rim}. In terms of the Fisher matrix analysis, this means that one should include priors that codify the CMB constraints on the acoustic horizon, so we have considered from \cite{Aghanim:2018eyx} the acoustic horizon $r_{drag} = 147.18 \pm 0.29 \, \mathrm{Mpc}$. Here we chose the standard procedure of including those priors as additional Fisher matrices that are added to the full Fisher matrix (for all parameters and all slices), before slicing and eventually inverting those matrices to find the constraints.

It is important to note that one may break degeneracies and improve measurements by measuring not only the BAO features, but also the shape of the power spectrum. However, since the shape measurements are much more sensitive to systematic errors than the pure BAO measurements \cite{Seo:2003pu,White:2008jy}, by isolating the former from the latter one obtains more robust constraints.
For that reason, it has become standard practice to first derive constraints from each redshift slice on $H(z)$ and $D_A(z)$, and then project those constraints into the cosmological parameters.

It has been pointed out that the smearing of the BAO scale caused by mode-coupling in the nonlinear regime can be partially undone (at least on large scales) by the procedure known as {\it reconstruction} \cite{Seo:2007ns}. For our dark energy constraints we assume that a simple, conservative reconstruction procedure has been applied to all datasets, which would lower the non-linear scale $\Sigma_0$ from 11$h^{-1}$Mpc to 6.5$h^{-1}$Mpc.

The procedure for extracting constraints from BAOs while isolating as much as possible the systematics from the unknown broad-band shape of the power spectrum and non-linear effects, has been well established \cite{Seo:2003pu}. We have followed this standard procedure, which in our case means that our basic (parent) Fisher matrices include not only the ``global'' degrees of freedom 
$\theta^{\rm glob} = \{\Omega_k,\Omega_b,\Omega_c,h,n_s\}$, but also ``local'' parameters, which are unknown on each redshift slice:
$\theta^{\rm loc} = \{ H(z), D_A(z), f \sigma_8 (z), b \sigma_8 (z), P_{shot}(z)\}$. If there are more than one tracer available on a given slice, there are as many bias factors in that slice.

After marginalizing against every other parameter in the parent Fisher matrix, we obtain constraints for the radial and angular-diameter distances on each redshift slice (for dark energy constraints we employed slices of $\Delta z = 0.2$, and rescaled DESI and Euclid parameters to match that choice). Finally, the Fisher matrices in terms of these parameters are used to derive constraints on the desired cosmological parameters -- in our case, $\{ \Omega_m, w_0, w_a \}$. This last step requires that we use the BAO scale, which is imposed in terms of a suitable prior derived from Planck data.

As mentioned earlier, our model for dark energy parametrizes the equation of state using two parameters, such that $w(a) = w_0 + w_a(1-a)$ \cite{Chevallier:2000qy,Linder:2002et}. The joint measurement of $w_0$ and $w_a$ has been the standard metric for comparing surveys in terms of their power to constrain dark energy \cite{2006astro.ph..9591A}.
In Fig. \ref{fig:w0wa} we compare the constraints on $w_0$ and $w_a$ for two areas of J-PAS, together with those for DESI and Euclid. In the top panel we show how the constraints improve as we include successive redshift slices, and in the bottom panel we show the added value of each successive slice for those constraints. In Fig. \ref{Figure_2} we plot 1$\sigma$ contour error for $w_0$ and $w_a$ using the information of all redshift bins. We summarize the marginalized errors for $w_0$ and $w_a$ in Table \ref{ta10b}.
\begin{figure*} %[h!]
 \begin{center}
 \includegraphics[width=0.46\textwidth]{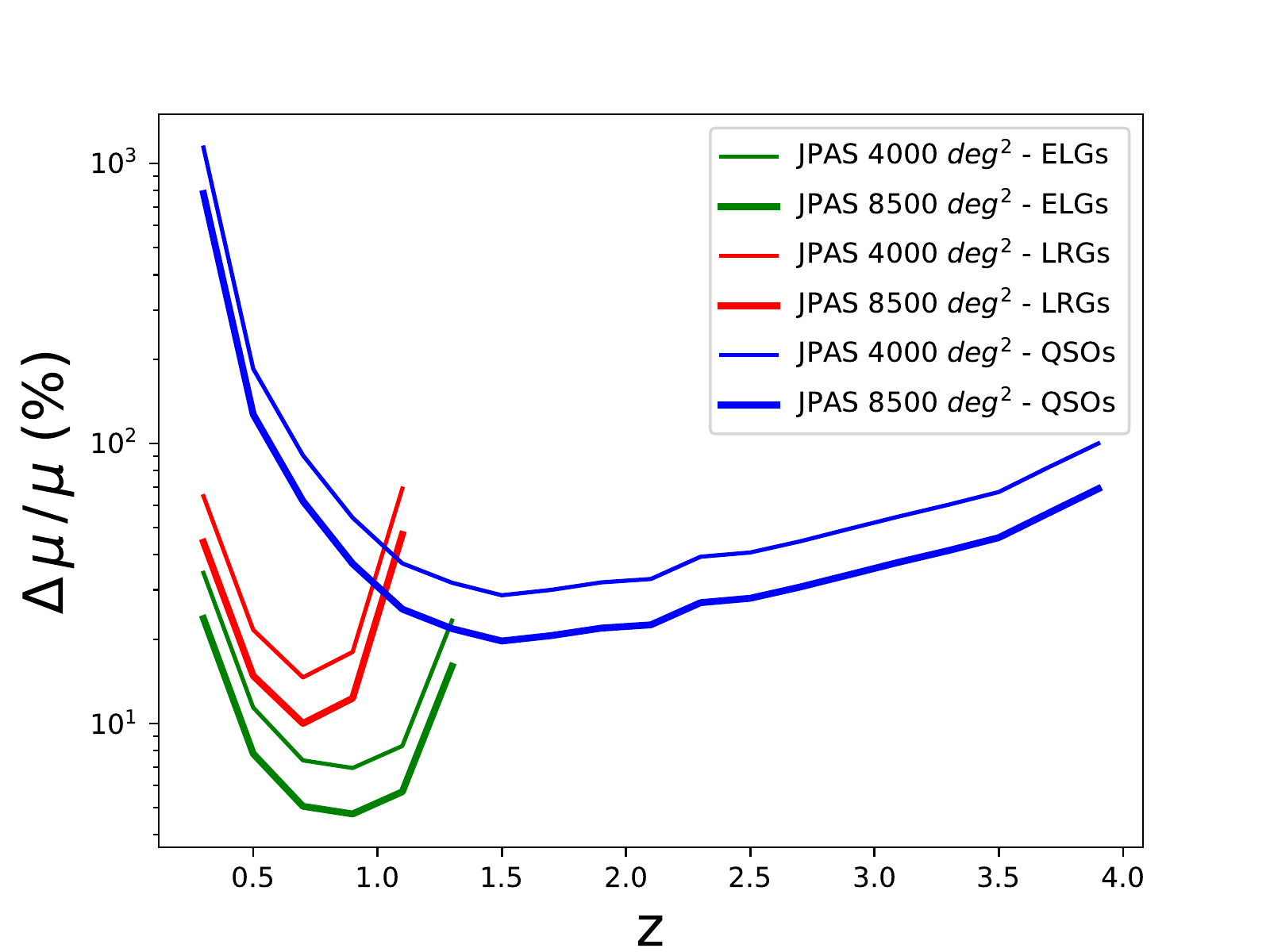} 
  \includegraphics[width=0.46\textwidth]{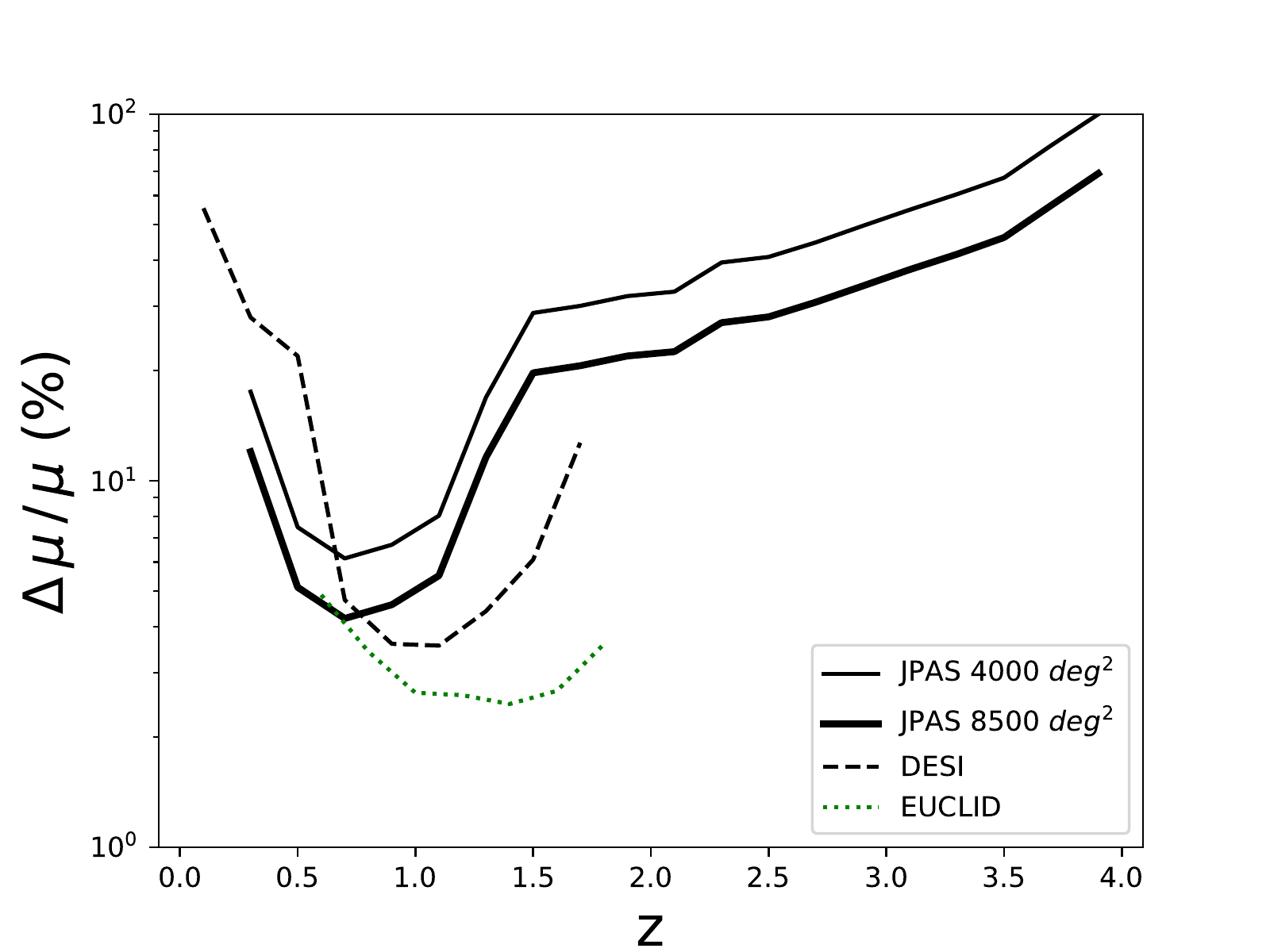}
  \end{center}
  \caption{Tomographic relative errors of $\mu$ for J-PAS with ELGs, LRGs and quasars (left). Tomographic relative errors of $\mu$ for J-PAS (ELGs+LRGs+QSOs), DESI (BGS+ELGs+LRGs+QSOs) and  Euclid (ELGs) using clustering information (right).}
 \label{Figure_3} 
 \end{figure*}

\begin{figure*} %[h!]
 \begin{center}
 \includegraphics[width=0.46\textwidth]{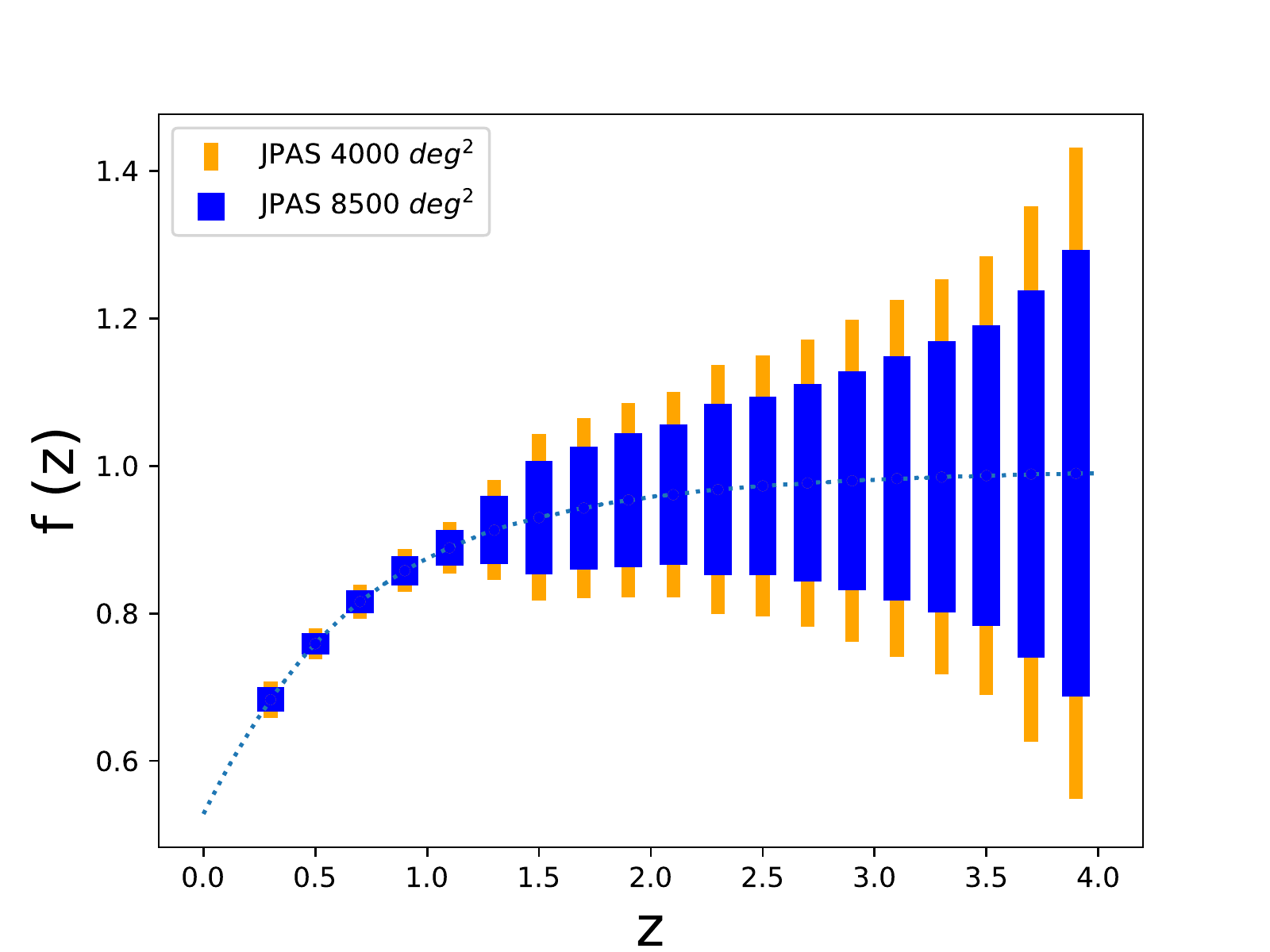}
 \includegraphics[width=0.46\textwidth]{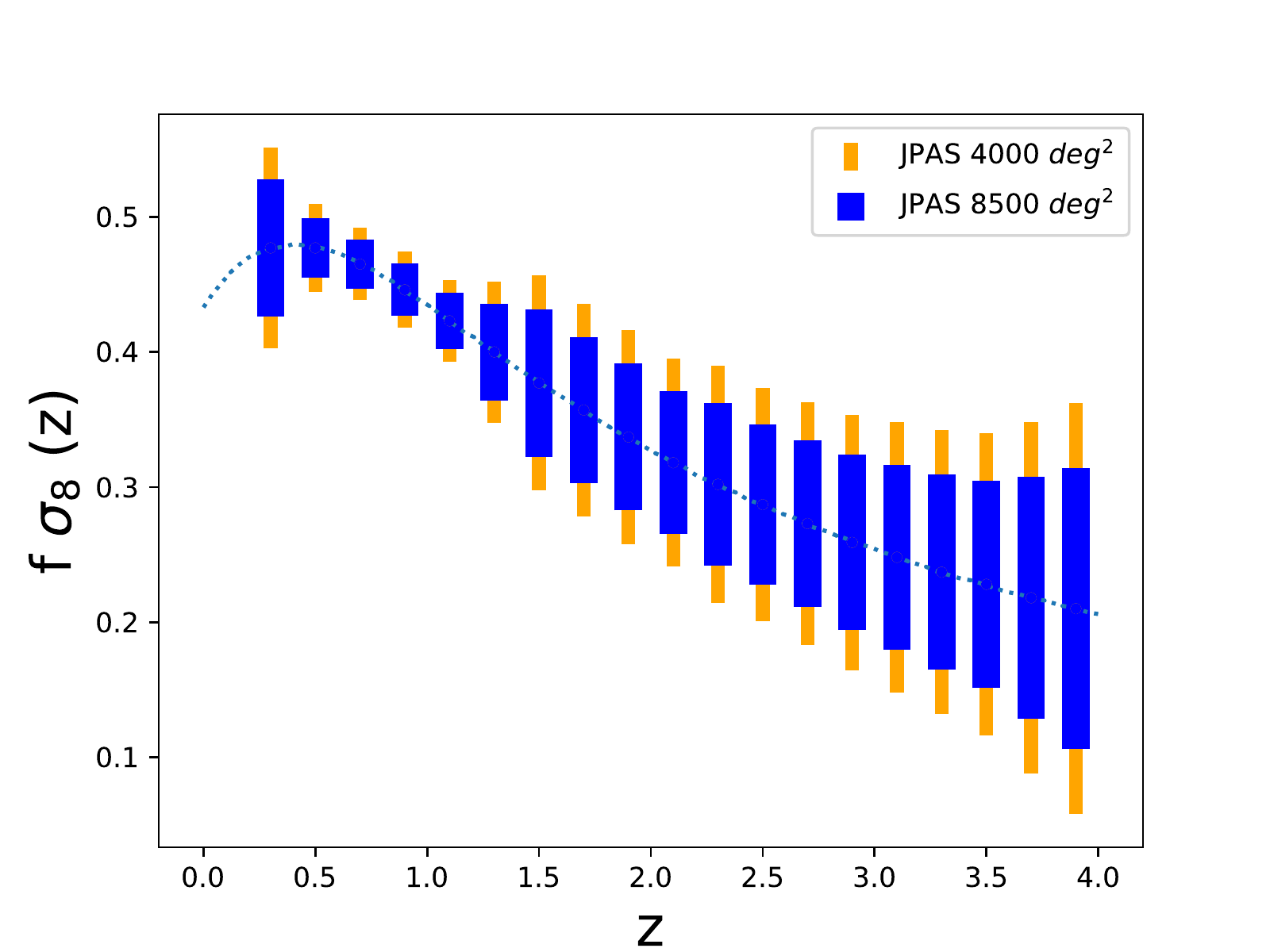}
 \end{center}
  \caption{Growth function and $f \sigma_8$ function for the fiducial cosmology with error bars for J-PAS 8500 and 4000 square degrees, using ELGs+LRGs+QSOs.}
 \label{Figure_1a} 
 \end{figure*}

\subsubsection{Modified Gravity}

For MG scenarios, we have the following independent parameters: $A_i$,  $R$ and  $E$ with $i$ denoting the different tracers. Because we have checked that marginalizing with respect to a non-Poissonian shot noise component has a minimal effect, for simplicity, we do not consider the shot noise term as a free parameter in this case. However, we are interested in obtaining errors for the effective Newton constant parameter $\mu$ and the growth function $f$.  Thus, we first consider as parameters the dimensionless quantities $A_i$, $R$ and $E$ for each redshift bin. Using the definitions of the $A_i$ and $R$ parameters we obtain for $\partial P_{i j}(k_{r},\hat{\mu}_{r},z_a) / \partial p_\alpha$,
\begin{subequations}
\begin{equation}\label{2.17}
\frac{\partial P_{i j}(k_{r},\hat{\mu}_{r},z_a)}{\partial A_{l}}=\left[ \frac{\delta_{l i}}{A_{i}+R\,\hat{\mu}^{2}} + \frac{\delta_{l j}}{A_{j}+R\,\hat{\mu}^{2}} \right] \, P_{i j},
\end{equation}
\begin{equation}\label{2.18}
\frac{\partial P_{i j}(k_{r},\hat{\mu}_{r},z_a)}{\partial R}= \left[ \frac{\hat{\mu}^{2}}{A_i+R\hat{\mu}^{2}} + \frac{\hat{\mu}^{2}}{A_j+R\hat{\mu}^{2}} \right] P_{i j},
\end{equation}
\begin{align}\label{2.19}
\frac{\partial P_{i j}(k_{r},\hat{\mu}_{r},z_a)}{\partial E} = &\left[\frac{1}{E}+2R\hat{\mu}^{2}(1-\hat{\mu}^{2})\,  \Xi \right.\nonumber \\
&\left.
+ \frac{2 \Delta z_{a}}{E^{2} \,H_{0} \chi(z_{a})} \right] P_{i j}\;
\end{align}
\end{subequations}
%
%\begin{eqnarray}\label{2.19}
%\frac{\partial P_{i j}(k_{r},\hat{\mu}_{r},z_a)}{\partial E} & =  & \left[\frac{1}{E}+\frac{2 \, \Delta z_{a}}{E^{2} \,H_{0} \, \chi(z_{a})}  + 2R\hat{\mu}^{2}(1-\hat{\mu}^{2}) \right.  
%  \left. \left( \frac{1}{A_i+R\,\hat{\mu}^{2}} + \frac{1}{A_j+R\,\hat{\mu}^{2}} \right) \nonumber \\ \left( \frac{1}{E}-\frac{\Delta z_{a}}{E^{2} \,H_{0} \, \chi(z_{a})} \right) \right] \, P_{i j}\; ,
%\end{eqnarray}
%
where 
$$
\Xi =  \left( \frac{1}{A_i+R\,\hat{\mu}^{2}} + \frac{1}{A_j+R\,\hat{\mu}^{2}} \right) \left( \frac{1}{E}-\frac{\Delta z_{a}}{E^{2} \,H_{0} \, \chi(z_{a})} \right)\;,
$$ 
and the length of the bin $\Delta z_{a}$ appears since we have discretized the integration in (\ref{2.8}) in order to calculate the derivative with respect to $E$. 
Following \cite{Amendola:2012ky}, in the calculation of $\partial P_{i j}(k_{r},\hat{\mu}_{r},z_a) / \partial E$ we do not consider the dependence of $P_{i j}(k_{r},\hat{\mu}_{r},z)$ on $E$ through $k$ since we do not know its explicit $k$ dependence in a model-independent way.

Once we have obtained the Fisher matrix for $ [\, A_i, \, R, \, E \,] $, we project first into $ [\, A_i, \, f, \, E \,] $, and then  to $ [\, A_i, \, \mu, \, E \,] $ using equations (\ref{25a}, \ref{3}) and the approximate analytic expression for $f=f(\mu,z)$ \cite{AparicioMaroto},
\begin{equation}
f(\mu,z)=\frac{1}{4} \left( \sqrt{1+24\,\mu}-1 \right) \, f_\Lambda(z),
\end{equation}
which is valid for time-independent $\mu$.  Thus, using (\ref{2.20}) we obtain the errors for $f$ and then those  for $\mu$. Forecasts for the  relative errors in $\mu$ and $f(z)$ in the different redshift bins  can be found in Table \ref{ta1} and in Table \ref{ta2} for J-PAS, in Table \ref{ta3} for DESI  and  in Table \ref{ta4} for Euclid. In Figure \ref{Figure_3} we plot these results for the three surveys. As we can see, ELGs provide the tightest constraints for J-PAS. Compared to Euclid or DESI, we find that J-PAS provides the best precision in the 
redshift range $z=0.3-0.6$. Notice this is also the case in the
$4000$ sq. deg. configuration.  This is mainly thanks to the large number
of expected ELG detection in that redshift range  which compensates the smaller fraction of sky of J-PAS as compared to other surveys.

\begin{figure*} %[h!]
 \begin{center}
 \includegraphics[width=0.46\textwidth]{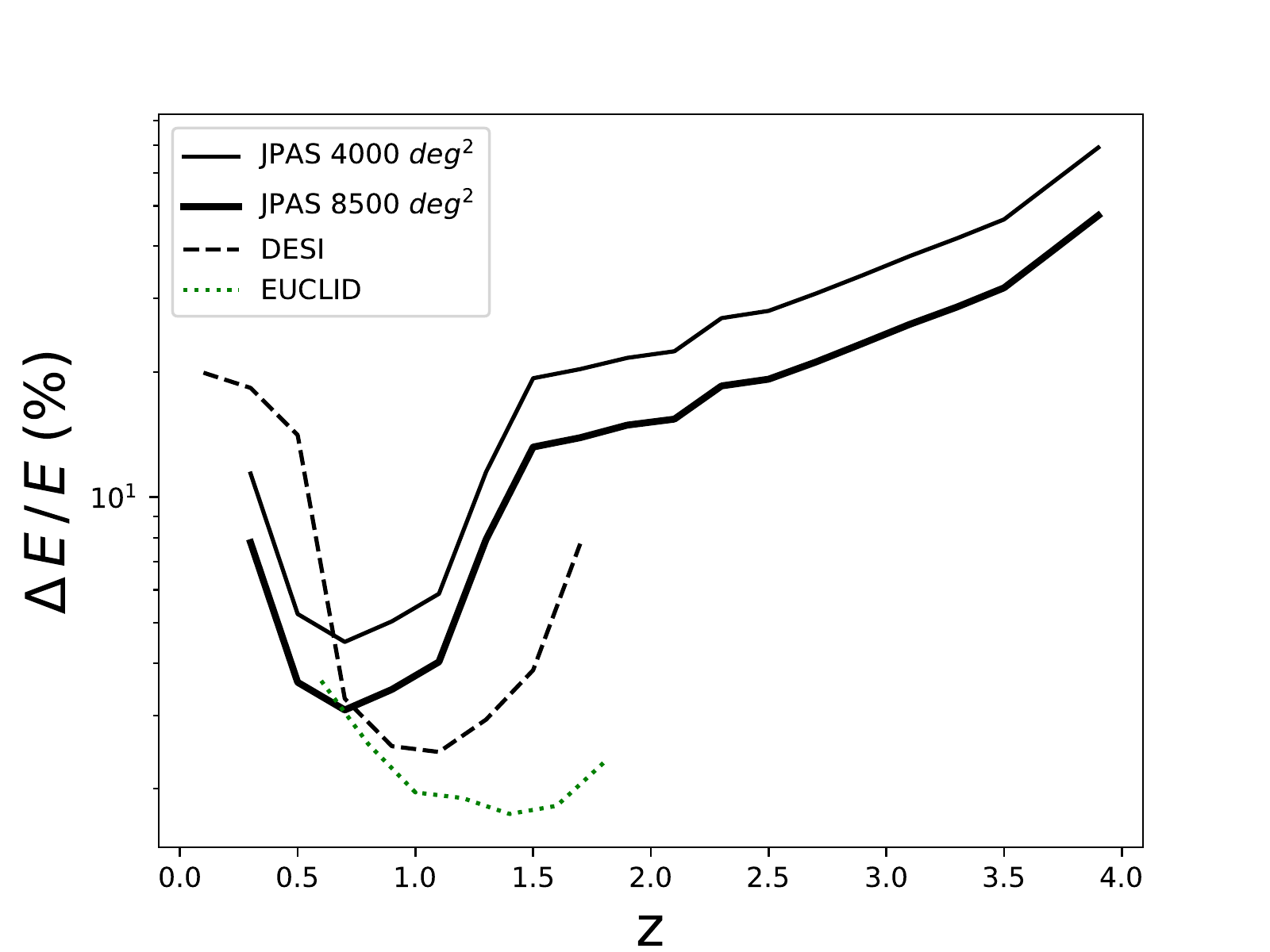}
 \includegraphics[width=0.46\textwidth]{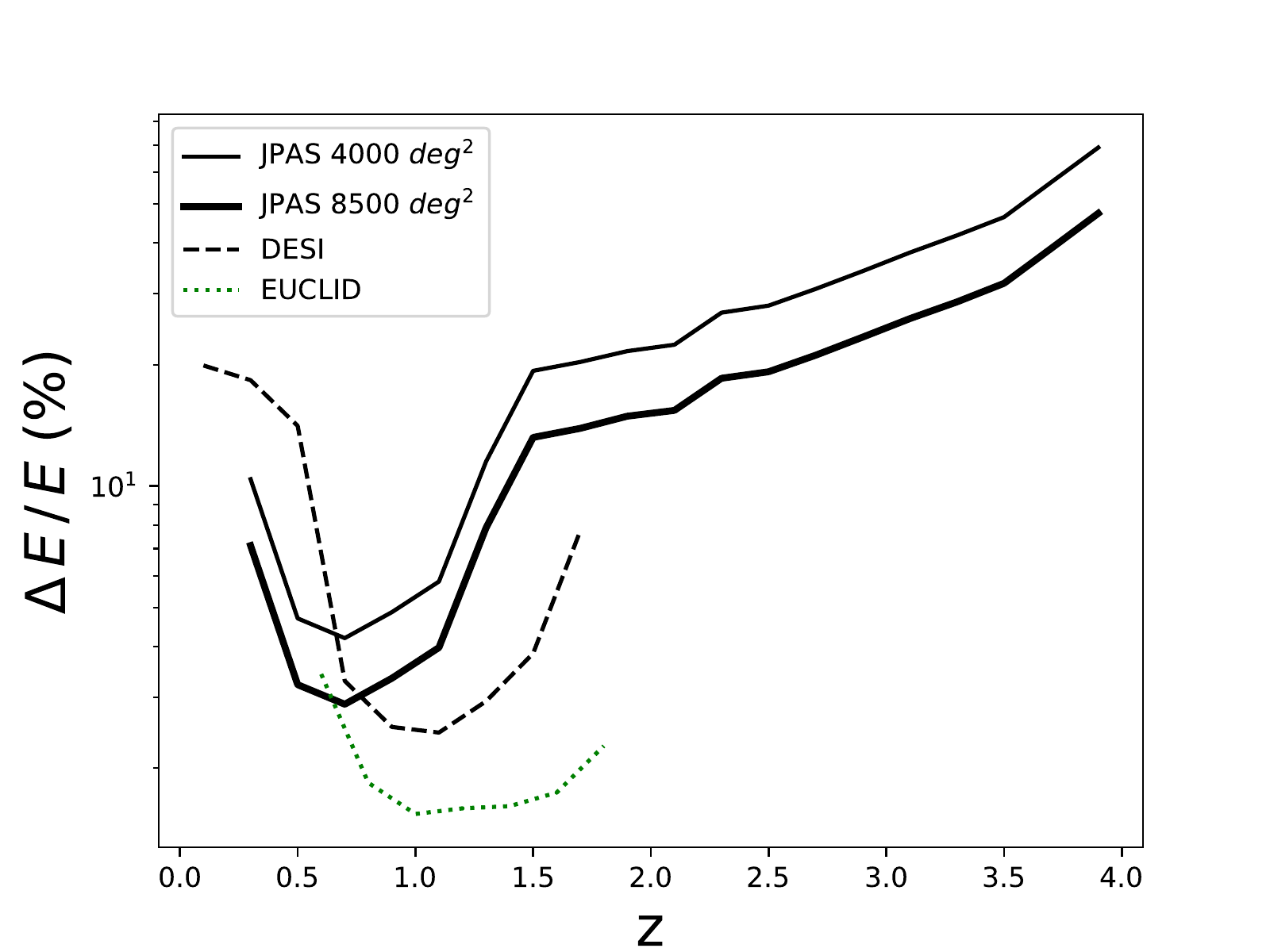}
  \end{center}
  \caption{Relative errors for $E(z)$ for J-PAS (ELGs+LRGs+QSOs), DESI (BGS+ELGs+LRGs+QSOs) and Euclid (ELGs) using clustering information (left panel), and using clustering and lensing information (right panel). In the case of DESI and J-PAS quasars, only clustering information is taken into account.  For lensing in J-PAS the redshift error is $\delta z=3 \%$.}
 \label{Figure_6} 
 \end{figure*}

In Figure \ref{Figure_1a} we show $f(z)$ and $f\sigma_8(z)$ with the expected error bars. 
Errors for $\mu$ in different $k$-bins are 
obtained using (\ref{2.22}) and can be found in Table \ref{ta1k} and in Figure \ref{Figure_1b} (left). We find that the best precision is obtained for scales around $k=0.1$ h/Mpc, which are slightly below
Euclid and DESI best scales.  Finally, in Figure \ref{Figure_6} (left) we show errors
for the Hubble dimensionless parameter $E(z)$ in the different redshift bins. Once more, J-PAS provides better precision below $z=0.6$, but also 
thanks to QSOs observation at higher redshifts, J-PAS will be able to 
measure the expansion rate in the practically unexplored region up to redshift $z=3.5$ with precision below $30\%$. 

%

%

%%%%%%%%%%%%%%%%%%%%%%%%%%%%%%%%%%%%%%%%%%%
\subsection{Weak Lensing} \label{WL}
%%%%%%%%%%%%%%%%%%%%%%%%%%%%%%%%%%%%%%%%%%

In this section, we obtain the errors on the $\eta$ parameter using weak lensing information. First, we compute the Fisher matrix for $[E,\,L]$ in each bin which has the following form,
\begin{eqnarray}
\left( \begin{array}{cccccc}
E_{1}E_{1} & E_{1}L_{1} & E_{1}E_{2} & E_{1}L_{2} & ... \\
L_{1}E_{1} & L_{1}L_{1} & L_{1}E_{2} & L_{1}L_{2} & ... \\
E_{2}E_{1} & E_{2}L_{1} & E_{2}E_{2} & E_{2}L_{2} & ... \\
L_{2}E_{1} & L_{2}L_{1} & L_{2}E_{2} & L_{2}L_{2} & ... \\
... & ... & ... & ... & ... \\
\end{array}\; \right)\;.
\\ \nonumber 
\end{eqnarray}

Then, we obtain the expressions for the derivatives of the convergence
power spectrum. The simplest case corresponds to the derivative with respect to $L$,
\begin{equation}\label{24}
\frac{\partial P_{ij}}{\partial L_{a}}=2H_{0} \, \frac{\Delta z_{a}}{E_{a}} \, K_{i} (z_{a}) K_{j} (z_{a}) \, L_{a} \, \hat{P}\left( \frac{\ell}{\pi \, \chi (z_{a})} \right).
\end{equation}
For the derivative with respect to $E$ we need the expression,
\begin{widetext}
\begin{eqnarray}\label{25}
\frac{\partial K_{i} (z_{b})}{\partial E_{a}}=\frac{3 (1+z_{b}) \Delta z_{a}}{2 E_{a}^{2}} \left[ - \hat{\theta} (z_{a}-z_{b}) \chi (z_{b})  \int_{z_{a}}^{\infty} \frac{n_{i} (z')}{\chi (z')^{2}} \, dz'  
 + \theta (z_{b}-z_{a}) \int_{z_{b}}^{\infty} \left(1-\frac{\chi (z_{b})}{\chi (z')} \right) \frac{n_{i} (z')}{\chi (z')}dz' \right],
\end{eqnarray}
\end{widetext}
where we have discretized the integration in equation  (\ref{2.8}) in the different bins and we have introduced Heaviside functions such that $\hat{\theta}(0)=0$ and $\theta(0)=1$.  Then the derivative with respect to $E$ reads,
\begin{eqnarray}\label{26}
&&\frac{\partial P_{ij}}{\partial E_{a}}=-H_{0}  \frac{\Delta z_{a}}{E_{a}^{2}}  K_{i} (z_{a}) K_{j} (z_{a})  L_{a}^{2}  \hat{P}\left( \frac{\ell}{\pi  \chi (z_{a})} \right)+\nonumber\\ &&
+H_{0} \sum_{b} \frac{\Delta z_{b}}{E_{b}}  \frac{\partial K_{i} (z_{b})}{\partial E_{a}}  K_{j} (z_{b}) \, L_{b}^{2} \hat{P}\left( \frac{\ell}{\pi  \chi(z_{b})} \right)+ \nonumber\\ &&
+H_{0} \sum_{b} \frac{\Delta z_{b}}{E_{b}} \frac{\partial K_{j} (z_{b})}{\partial E_{a}}  K_{i} (z_{b}) \, L_{b}^{2} \hat{P}\left( \frac{\ell}{\pi  \chi(z_{b})} \right). 
\end{eqnarray}

\begin{figure*} %[h!]
 \begin{center} 
  \includegraphics[width=0.32\textwidth]{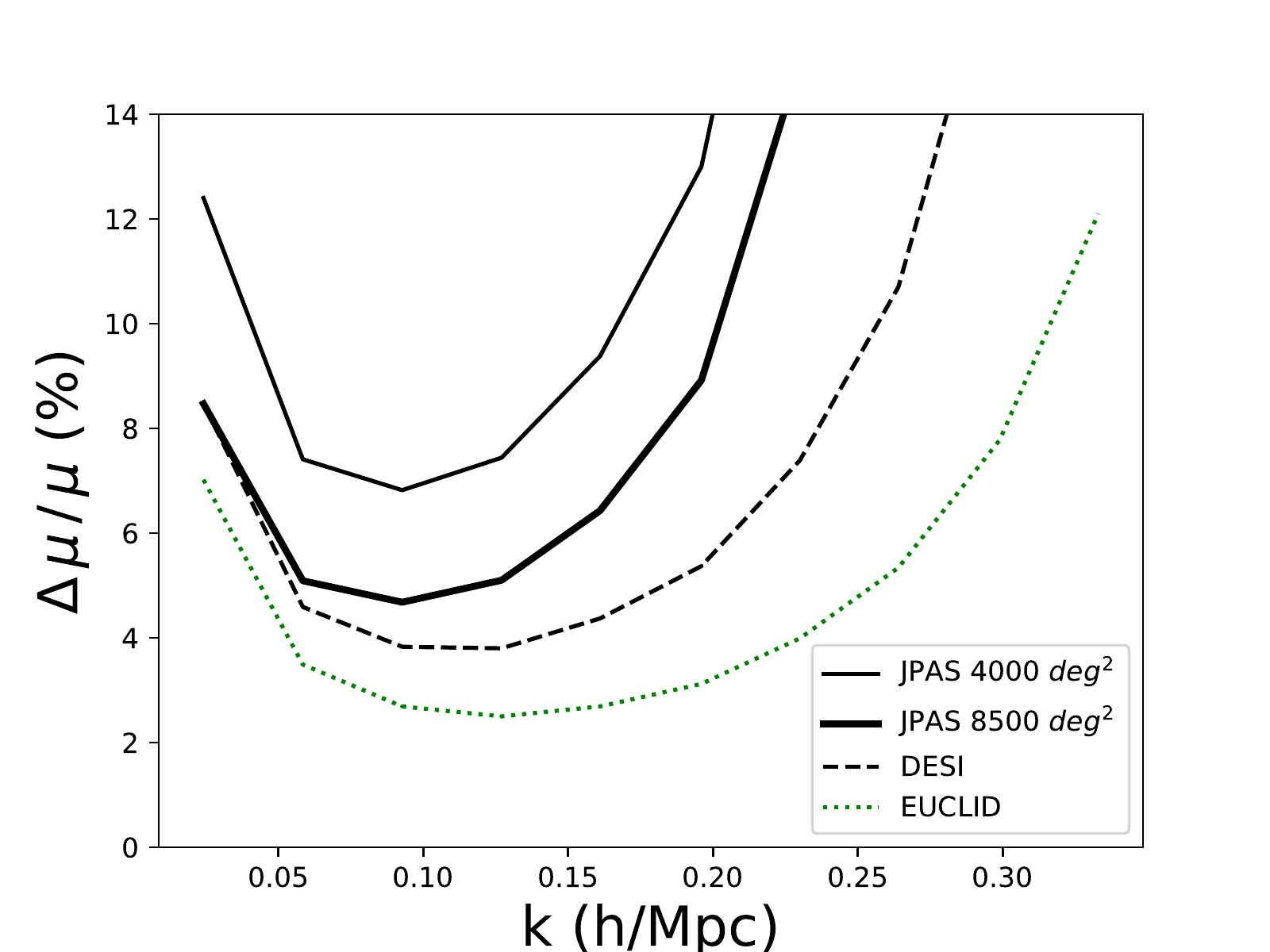}
   \includegraphics[width=0.32\textwidth]{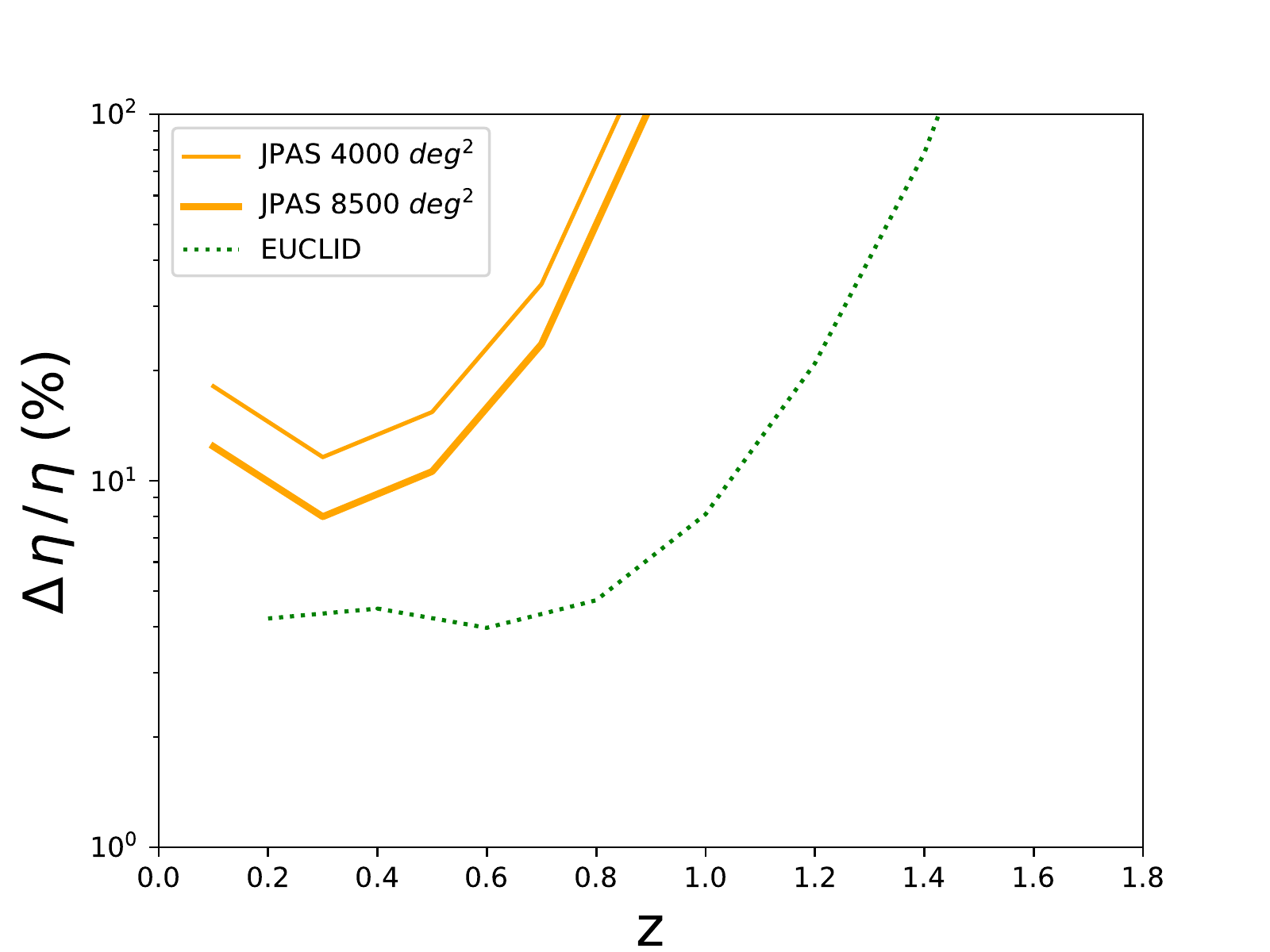} 
    \includegraphics[width=0.32\textwidth]{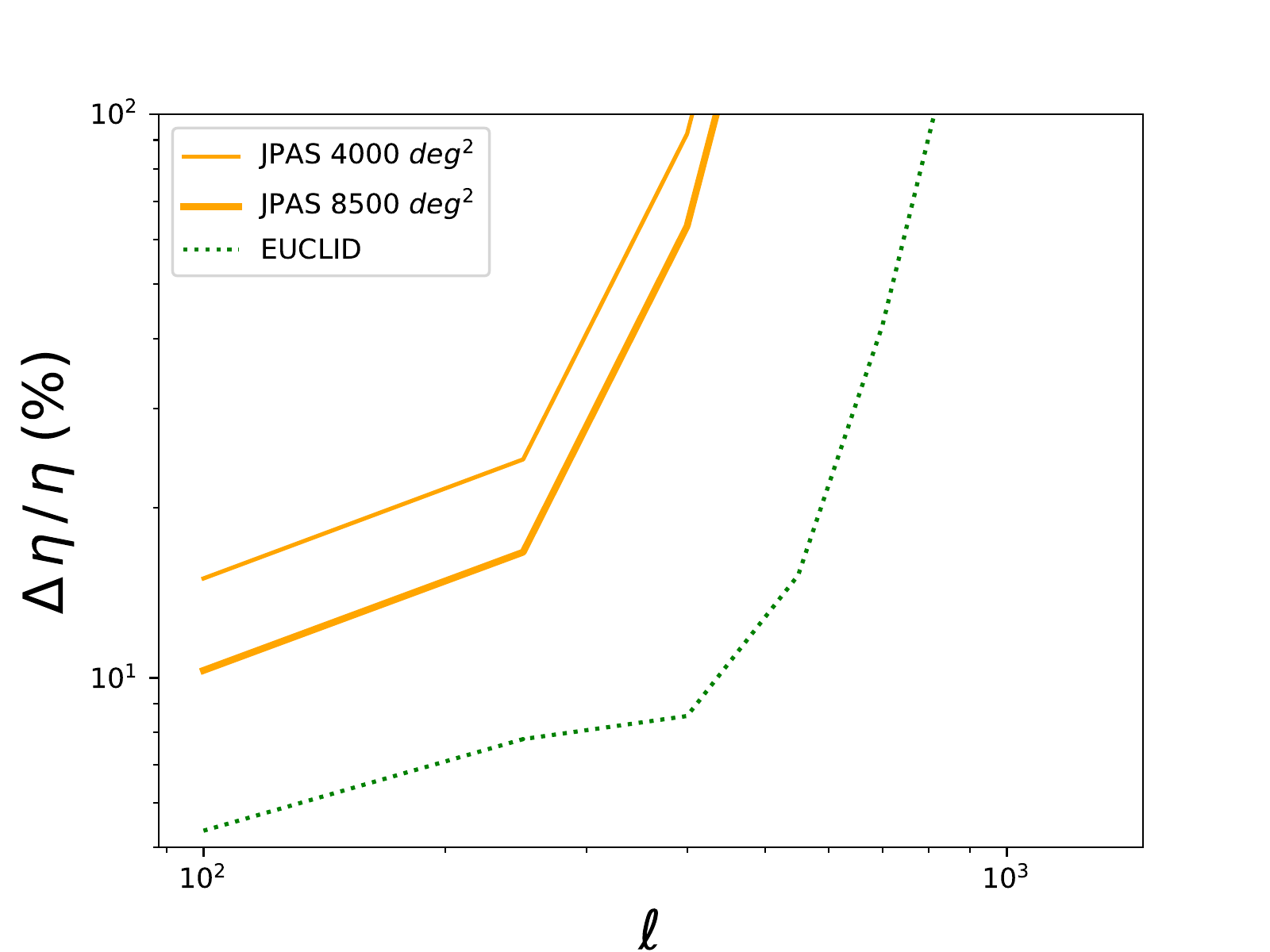} 
  \end{center}
  \caption{{\it{left)}} Relative errors of $\mu(k)$ for J-PAS (ELGs+LRGs+QSOs), DESI (BGS+ELGs+LRGs+QSOs) and  Euclid (ELGs) using clustering information. {\it{middle)}} Tomographic relative errors of $\eta$ for J-PAS (ELGs+LRGs) and Euclid (ELGs) using lensing information. {\it{right)}} Relative errors of $\eta (\ell)$ for J-PAS (ELGs+LRGs) and Euclid (ELGs) using lensing information.}
 \label{Figure_1b} 
 \end{figure*}
%

% \begin{figure*} %[h!]
%  \begin{center}
%  \includegraphics[width=0.49\textwidth]{GF.pdf}
%  \includegraphics[width=0.49\textwidth]{f_s8_err.pdf}
%   \end{center}
%   \caption{Growth function and $f \sigma_8$ function for the fiducial cosmology with error bars for J-PAS 8500 and 4000 square degrees, using ELGs+LRGs+QSOs.}
%  \label{Figure_1a} 
%  \end{figure*}
%

\begin{figure*} %[h!]
 \begin{center}
 \includegraphics[width=0.46\textwidth]{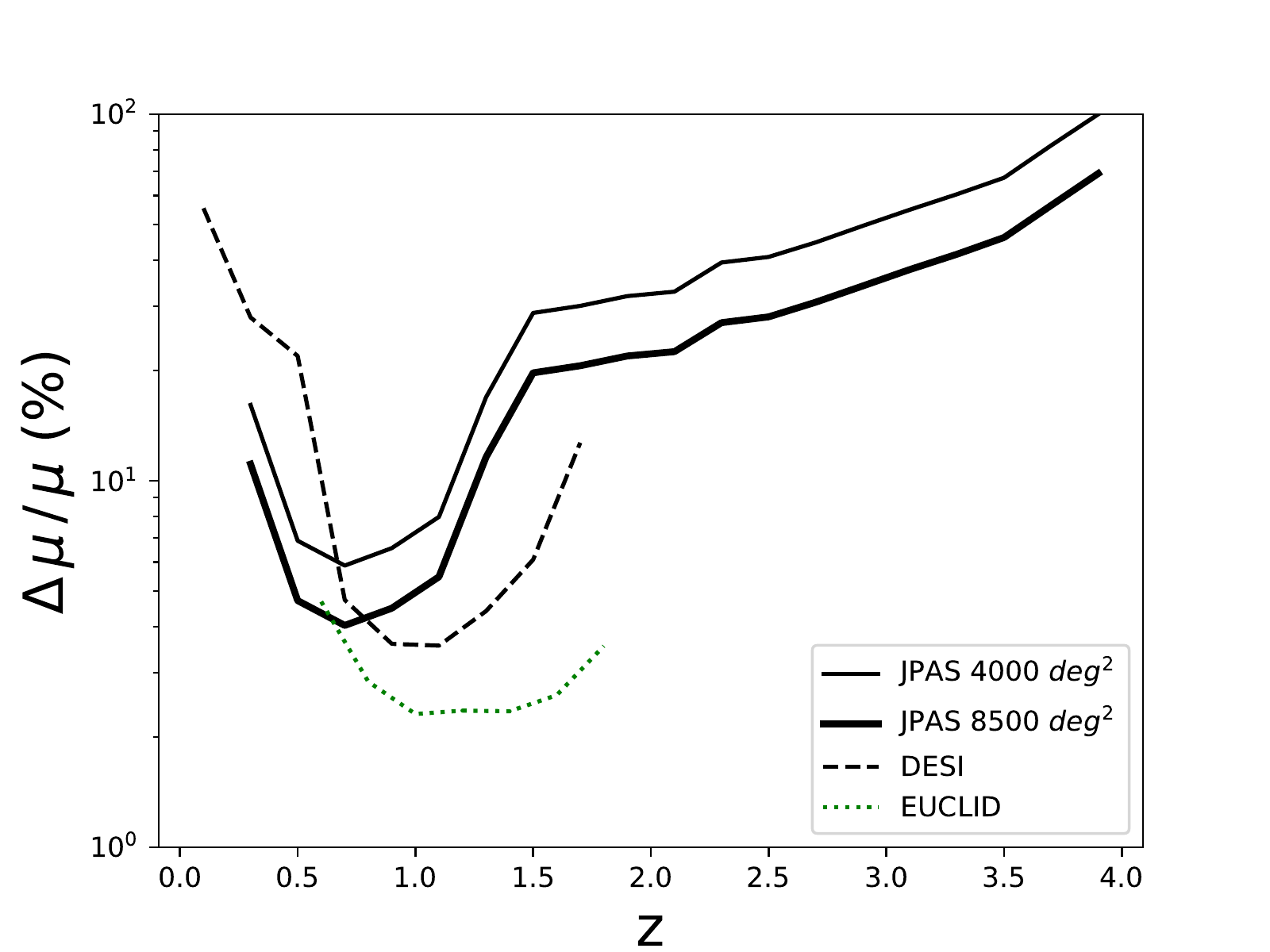} 
  \includegraphics[width=0.46\textwidth]{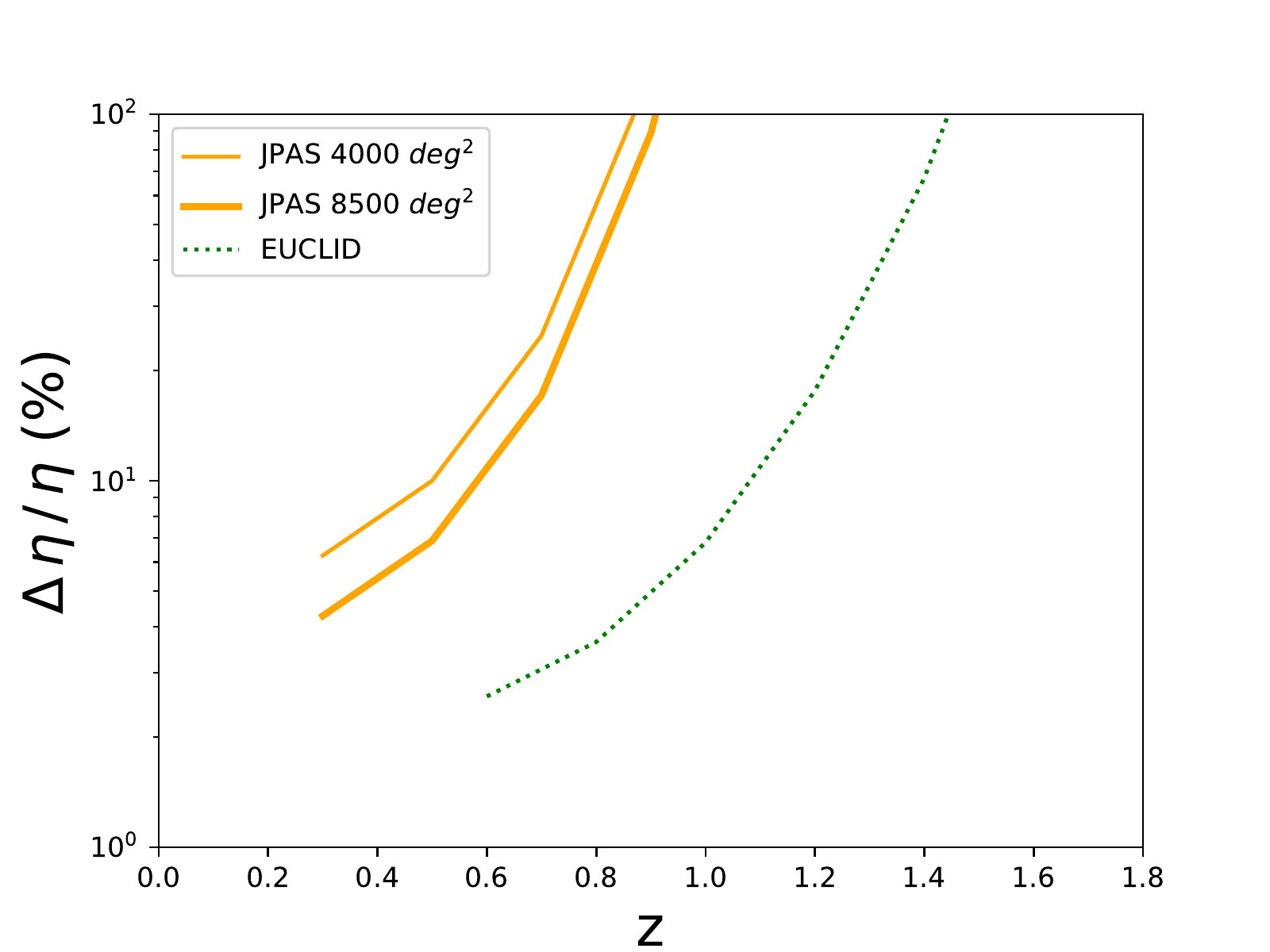}
  \end{center}
  \caption{From left to right, tomographic relative errors for $\mu$ and $\eta$ for J-PAS (ELGs+LRGs+QSOs), DESI (BGS+ELGs+LRGs+QSOs) and Euclid (ELGs) using clustering and lensing information. In the case of DESI and J-PAS quasars only clustering information is taken into account.  For lensing in J-PAS the redshift error is $\delta z=3 \%$.}
 \label{Figure_5} 
 \end{figure*}
As in the clustering case, we have not considered derivatives of  $\hat{P}(k)$.

Now it is necessary to change the initial parameters $[E, L]$ to the new ones $[E, \eta]$. Using (\ref{16}) we obtain $\frac{\partial \eta}{\partial L} = \frac{2}{L}$ and $\frac{\partial \eta}{\partial E} = 0$. For time-independent parameters, we show in Table \ref{ta6} and in Figure \ref{Figure_1b} (middle)  the relative errors in $\eta$ for the different redshift bins for J-PAS and Euclid. 
Again, J-PAS provides the best errors in the range $z=0.3-0.6$. 
In order to obtain the errors of $\eta$ in different $\ell$-bins we compute the Fisher matrix (\ref{21b}). We first change from $[E, L]$ to $[E, \eta]$ in each redshift bin and then sum the information of $\eta$ for the different redshift bins. The corresponding errors can be found in Table \ref{ta7} for J-PAS and Euclid as well as in Figure \ref{Figure_1b} (right).   

%
% \begin{figure*} %[h!]
%  \begin{center}
%  \includegraphics[width=0.7\textwidth]{Err_eta1.pdf} 
%   \end{center}
%   \caption{Relative errors of $\eta (z)$ for J-PAS (ELGs+LRGs) and Euclid (ELGs) using lensing information.}
%  \label{Figure_2} 
%  \end{figure*}
% %
% 
% %
% \begin{figure*} %[h!]
%  \begin{center}
%  \includegraphics[width=0.7\textwidth]{Error_eta_l.pdf} 
%   \end{center}
%   \caption{Relative errors of $\eta (\ell)$ for J-PAS (ELGs+LRGs) and Euclid (ELGs) using lensing information.}
%  \label{Figure_2b} 
%  \end{figure*}
% %

\begin{figure*}%[h!]
 \begin{center}
 \includegraphics[height=5.3cm]{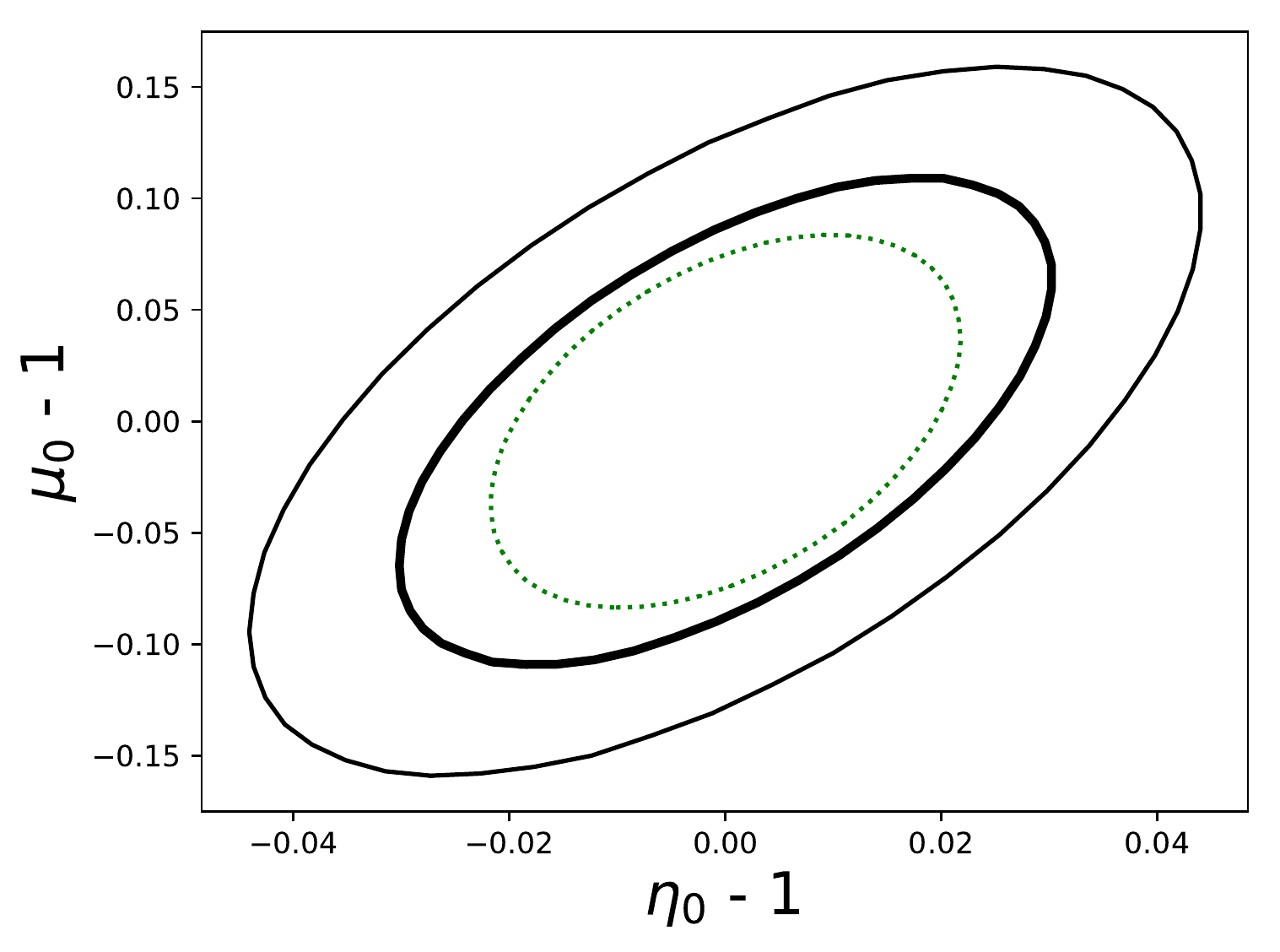} 
  \includegraphics[height=5.3cm]{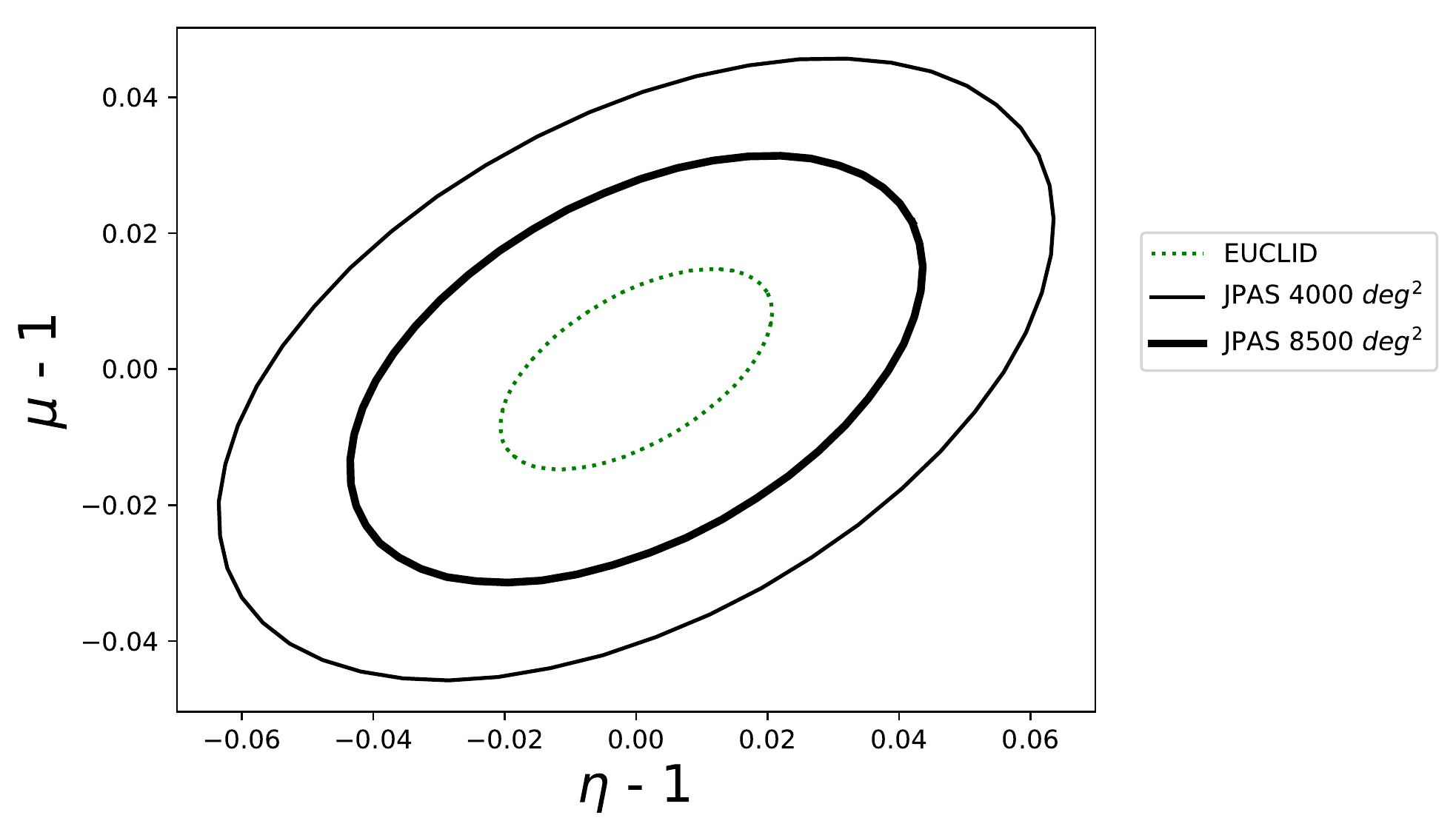}
  \end{center}
  \caption{(Left panel) 1$\sigma$ contour error for $\mu_0$ and $\eta_0$ defined in  (\ref{31}) and (\ref{32}), and (right panel) for constant $\mu$ and $\eta$. All in J-PAS (ELGs+LRGs+QSOs) and Euclid (ELGs) surveys combining clustering and lensing information, for 8500 $\mathrm{deg}^2$ and 4000 $\mathrm{deg}^2$.}
 \label{Figure_7} 
 \end{figure*}

%%%%%%%%%%%%%%%%%%%%%%%%%%%%%%%%%%%%%%%%%%%
\subsection{Clustering + Weak Lensing} \label{C+WL}
%%%%%%%%%%%%%%%%%%%%%%%%%%%%%%%%%%%%%%%%%%

Finally, in this section we analyze the case in which information from clustering and lensing is combined. We first take the Fisher matrix of parameters $[A_i, \mu, E]$ for clustering and $[E, \eta]$ for weak lensing and build the full matrix with parameters $[A_i, \mu, E, \eta]$. This matrix has the form,
\[ \left( \begin{array}{ccccc}
A_{1}A_{1} & A_{1} \mu_{1} & 0 & A_{1}E_{1} & ...  \\
\mu_{1} A & \mu_{1} \mu_{1} & 0 & \mu_{1} E_{1} & ... \\
0 & 0 & \eta_{1} \eta_{1} & \eta_{1} E_{1} & ...  \\
E_{1}A_{1} & E_{1} \mu_{1} & E_{1} \eta_{1} & E_{1}E_{1} & ... \\
... & ... & ... & ... & ... \end{array} \right)\]
\normalsize
where $EE$ is the sum of terms $EE$ for clustering and lensing. By inverting this Fisher matrix, we obtain the errors for $\mu$ and $\eta$. These results are shown in Table \ref{ta8} for J-PAS and in Table \ref{ta9} for Euclid. Finally, Figure \ref{Figure_5} compares the sensitivity of both surveys for
time-independent $\mu$ and $\eta$ in the different redshift bins. For completeness, we also show the same comparison for the function $E(z)$ in Figure \ref{Figure_6}. As we can see, the combination of clustering and lensing information improves the sensitivity in around a $10\%$ for all the parameters.
 We sum all the information in the whole redshift range for $\mu$ and $\eta$ and plot their error ellipses in the  right panel of Figure \ref{Figure_7}. These results are summarized in Table \ref{ta10}.

So far we have limited ourselves to time-independent $\mu$ and
$\eta$ parameters. For scale-independent parameters, we consider the case in (\ref{31}) and (\ref{32}). Using the analytical fitting function for this particular expressions obtained in  \cite{AparicioMaroto}, we  obtain errors for $\mu_0$ and $\eta_0$ with  fiducial values $\mu_0 = \eta_0 = 1$. We plot on the left panel of Figure \ref{Figure_7} error ellipses for $\mu_0$ and $\eta_0$, and we summarize these errors in Table \ref{ta10}.

%%%%%%%%%%%%%%%%%%%%%%%%%%%%%%%%%%%%%%%%%%%%%%%%%%%%%%%%%%%%%%%%%%%%%%%%%%%%%%%%%%%%
\section{Discussion and Conclusion}
%%%%%%%%%%%%%%%%%%%%%%%%%%%%%%%%%%%%%%%%%%%%%%%%%%%%%%%%%%%%%%%%%%%%%%%%%%%%%%%%%%%%

Over the past years, cosmological observations have been used not only to constrain  models within the context of GR but also the theory of gravity itself (see e.g. \cite{Okumura:2015lvp}). In general, MG theories introduce changes in the Poisson equation which relate the density perturbations $\delta$ with the gravitational potential $\Psi$, thus modifying the amplitude and evolution of the growth of cosmological perturbations. Furthermore, gravitational lensing is directly sensitive to the growth of dark matter perturbations -- in contrast with measurements based on galaxies, neutral hydrogen or any other baryonic tracer. These theories, therefore, also introduce modifications in the equation that determines the lensing potential and controls the motion of photons. Thus, observations of the distribution of matter on large scales at different redshifts, and of the weak lensing generated by those structures, provide a new suite of tests of GR on cosmological scales \cite{Tsujikawa:2012hv, Huterer:2013xky, Joyce:2014kja}.

In this work we have investigated the ability of the J-PAS survey to constrain dark energy and MG cosmologies using both the J-PAS information on the galaxy power spectra for different dark matter tracers, with baryon acoustic oscillations and redshift-space distortions, as well as the weak lensing information by considering the convergence power spectrum. Our analysis considers phenomenological  parameterization of dark energy and modified gravity models, as discussed in Sec.~\ref{Sec:MGmodels}. 

Following \cite{Amendola:2012ky}, we have adopted a model-independent parameterization of the power spectra of clustering and weak lensing. This parameterization considers all the free and independent parameters that are needed to describe such power spectra in the linear regime. In this analysis, we have fixed the initial dark matter power spectrum $\hat P(k)$ to the fiducial model, corresponding to a flat $\Lambda$CDM cosmology. As mentioned above, rather than focusing on specific dark energy or MG theories, we have considered a phenomenological  approach described in terms of 
a set of parameters that can be contrasted with observations. 
Thus, in the dark energy case, the widely used $(w_0,w_a)$ CPL parameterization has been assumed. 
For MG theories, two cases have been considered. First, for time-independent 
$\mu$ and $\eta$, we have performed both a tomographic redshift bin analysis and an analysis in $k$-bins. By summing over all the redshift range we have obtained the best errors for the modified gravity parameters. Second, for scale-independent parameters, we have considered the particular parameterization in terms of $\mu_0$ and $\eta_0$ (\ref{31}-\ref{32}) usually employed in the literature.

J-PAS will be able to measure different tracers, e.g. LRG, ELG and QS. In order to contextualize the J-PAS measurements, we have performed the same Fisher analysis for DESI and Euclid surveys. In the case of DESI, in addition to LRGs, ELGs and QSOs, a bright galaxy sample (BGS) will be also measured at low redshifts, while Euclid will measure only ELGs. In the dark energy analysis, we have found that
J-PAS will measure $w_0$ with precision below 6$\%$ that can be compared with the $4.5\%$ for DESI and $3\%$ for Euclid.
The absolute error in $w_a$ is found to be below $0.24$ for J-PAS, $0.19$ for DESI and $0.13$ for Euclid. 
From the tomographic analysis, we find that using the clustering information alone,
J-PAS will allow to measure the expansion rate $H(z)$ with precision $3\%$ in the best redshift bin ($z=0.7$) and the $\mu$ parameter with a precision  around 5$\%$ in the best redshift bin. From lensing alone, J-PAS will be able to measure $\eta$ with a precision around $8 \%$ in the best redshift bin. 
The combination of clustering and lensing will allow to improve the precision in $\mu$ down to $4 \%$ in the best bin. Considering the information  in the whole redshift range, we have found that J-PAS will be able to measure time-independent $\mu$ and $\eta$ with precision better than $3\%$ for both parameters. For $\mu_0$ and $\eta_0$ we have obtained errors of $10 \%$ and $5 \%$, respectively. 

When compared to future spectroscopic surveys such as DESI or spectroscopic and photometric ones such as Euclid, we have shown that from clustering and lensing information, J-PAS will have the best errors for redshifts between $z=0.3-0.6$, thanks to the large number of ELGs detectable in that redshift range. Note  also that 
thanks to QSOs observation at higher redshifts, J-PAS will be able to 
measure the expansion rate and MG parameters in the practically unexplored region up to redshift $z=3.5$ with precision below $30\%$.

 In the whole redshift range, the J-PAS precision in both $\mu$ and $\eta$ will be a factor 1.5-2 below Euclid in their respective best bins. For the (time-dependent) $\mu_0$ - $\eta_0$ parameterization (\ref{31}-\ref{32}), we have shown that J-PAS is closer to Euclid than in the constant case. This is due to the fact that low-redshift measurements are more sensitive to $\mu_0$ and $\eta_0$ than high-redshift ones, such that at low redshift J-PAS precision surpasses that of Euclid.

Finally, it is worth mentioning that by increasing  the precision in the determination of the  dimensionless Hubble parameter using e.g. the J-PAS sample of type Ia supernovae, and taking into account information from the non-linear power spectra, it can be expected that the 
sensitivity to the $\mu$ and $\eta$ parameters will increase. Additionally, considering the cross correlation between galaxy distribution and galaxy shapes will also allow to improve the 
precision of J-PAS in the determination of dark energy and MG parameters. 

% the type Ia supernovae data of J-PAS, which would increase the accuracy in the dimensionless Hubble parameter, increasing the accuracy in $\mu$ and $\eta$. Moreover, we have only considered the linear regime, but considering the non-linear power spectra it is expected that the constraints on the parameters $\mu$ and $\eta$ would increase. 

%%%%%%%%%%%%%%%%%%%%%%%%%%%%%%%%%%%%%%%%%%%%%%%%%%%%%%%%%%%%%%%%%%%%%%%%%%%%%%%%%%%%%%%%%%%%%%%%%%%%%%%%%%

\vspace{0.5cm}

\section*{Acknowledgements} 

We are thankful to our colleagues of J-PAS Theory Working Group for useful discussions and to Ricardo Landim for his comments.  MAR and ALM acknowledge support from MINECO (Spain) project  FIS2016-78859-P(AEI/FEDER, UE), Red Consolider MultiDark FPA2017-90566-REDC and UCM predoctoral grant. JSA acknowledges support from FAPERJ grant no. E-26/203.024/2017, CNPq grant no. 310790/2014-0 and 400471/2014-0 and the Financiadora  de  Estudos  e  Projetos  -  FINEP  grants  REF.  1217/13  -  01.13.0279.00  and  REF0859/10  -  01.10.0663.00. SC acknowledges support from CNPq grant no. 307467/2017-1 and 420641/2018-1.

This paper has gone through internal review by the J-PAS collaboration.
Funding for the J-PAS Project has been provided by the
Governments of España and Aragón through the Fondo de Inversión
de Teruel, European FEDER funding and the MINECO and
by the Brazilian agencies FINEP, FAPESP, FAPERJ and by the National
Observatory of Brazil.

\bibliography{JPasMG}

\clearpage
%\printbibliography
%\clearpage

\begin{widetext}

%%%%%%%%%%%%%%%%%%%%%%%%%%%%%%%%%%%%%%%%%%%%%%%%%%%%%%%%%%%%%%%%%%%%%%%%%%%%%%%%%%%%
\section{Appendix A: data tables}\label{app}
%%%%%%%%%%%%%%%%%%%%%%%%%%%%%%%%%%%%%%%%%%%%%%%%%%%%%%%%%%%%%%%%%%%%%%%%%%%%%%%%%%%%

\begin{table*}[htbp]
\begin{center}
\begin{tabular}{|c|c|c|c|}
\hline
\multicolumn{4}{|c|}{J-PAS} \\ \hline
\hline
$z$ & $LRG$ & $ELG$ & $QSO$ \\
\hline \hline
0.3 & 226.6 & 2958.6 & 0.45 \\ \hline
0.5 & 156.3 & 1181.1 & 1.14 \\ \hline
0.7 & 68.8  & 502.1  & 1.61 \\ \hline
0.9 & 12.0  & 138.0  & 2.27 \\ \hline
1.1 & 0.9   & 41.2   & 2.86 \\ \hline
1.3 & 0     & 6.7    & 3.60 \\ \hline
1.5 & 0     & 0      & 3.60 \\ \hline
1.7 & 0     & 0      & 3.21 \\ \hline
1.9 & 0     & 0      & 2.86 \\ \hline
2.1 & 0     & 0      & 2.55 \\ \hline
2.3 & 0     & 0      & 2.27 \\ \hline
2.5 & 0     & 0      & 2.03 \\ \hline
2.7 & 0     & 0      & 1.81 \\ \hline
2.9 & 0     & 0      & 1.61 \\ \hline
3.1 & 0     & 0      & 1.43 \\ \hline
3.3 & 0     & 0      & 1.28 \\ \hline
3.5 & 0     & 0      & 1.14 \\ \hline
3.7 & 0     & 0      & 0.91 \\ \hline
3.9 & 0     & 0      & 0.72 \\ \hline
\end{tabular}\hspace{1in}\begin{tabular}{|c|c|c|c|c|}
\hline
\multicolumn{5}{|c|}{DESI} \\ \hline
\hline
$z$ & $BGS$ & $LRG$ & $ELG$ & $QSO$ \\
\hline \hline
0.1 & 2240 & 0    & 0    & 0      \\ \hline
0.3 & 240  & 0    & 0    & 0      \\ \hline
0.5 & 6.3  & 0    & 0    & 0      \\ \hline
0.7 & 0    & 48.7 & 69.1 & 2.75   \\ \hline
0.9 & 0    & 19.1 & 81.9 & 2.60   \\ \hline
1.1 & 0    & 1.18 & 47.7 & 2.55   \\ \hline
1.3 & 0    & 0    & 28.2 & 2.50   \\ \hline
1.5 & 0    & 0    & 11.2 & 2.40   \\ \hline
1.7 & 0    & 0    & 1.68 & 2.30   \\ \hline
\end{tabular}\hspace{1in}\begin{tabular}{|c|c|}
\hline
\multicolumn{2}{|c|}{Euclid} \\ \hline
\hline
$z$ & $ELG$ \\
\hline \hline
0.6 & 356 \\ \hline
0.8 & 242 \\ \hline
1.0 & 181 \\ \hline
1.2 & 144 \\ \hline
1.4 & 99  \\ \hline
1.6 & 66  \\ \hline
1.8 & 33  \\ \hline
\end{tabular}
\caption{In left panel: redshift bins and densities of luminous red galaxies, emission line galaxies and quasars for J-PAS. In center panel: redshift bins and densities of bright galaxies, luminous red galaxies, emission line galaxies and quasars for DESI. In right panel: redshift bins and densities of emission line galaxies for Euclid. Galaxy densities in units of $10^{-5}$ $\mathrm{h^3 \, Mpc^{-3}}$.}
\label{taJPASDESI}
\end{center}
\end{table*}

\begin{table*}[htbp]
\centering
\begin{tabular}{|c|c|c|}
\hline
Survey & $\Delta w_0$ & $\Delta w_a$   \\
\hline \hline
Euclid       & 0.029 & 0.128  \\ \hline
DESI         & 0.045 & 0.186  \\ \hline
J-PAS $8500$ & 0.058 & 0.238  \\ \hline
J-PAS $4000$ & 0.079 & 0.316  \\ \hline
\end{tabular}
\caption{Absolute errors for $w_0$ and $w_a$ for Euclid, DESI, and JPAS (with 8500 and 4000 square degrees), considering clustering information.}
\label{ta10b}
\end{table*}
\begin{table*}[htbp]
\begin{center}
\begin{tabular}{|c|c|c|c|c|c|}
\hline
\multicolumn{6}{|c|}{DESI clustering} \\ \hline
\hline
$z$  & $\mu$ & $\Delta \mu / \mu (\%)$ & $f$ & $\Delta f$ & $\Delta f / f (\%)$ \\
\hline \hline
0.1 & 1 & 55.4 & 0.585 & 0.085 & 14.5  \\ \hline
0.3 & 1 & 27.9 & 0.683 & 0.037 & 5.47  \\ \hline
0.5 & 1 & 21.9 & 0.759 & 0.048 & 6.32  \\ \hline
0.7 & 1 & 4.73 & 0.816 & 0.016 & 1.96  \\ \hline
0.9 & 1 & 3.59 & 0.858 & 0.014 & 1.62  \\ \hline
1.1 & 1 & 3.55 & 0.889 & 0.014 & 1.58  \\ \hline
1.3 & 1 & 4.41 & 0.913 & 0.017 & 1.87  \\ \hline
1.5 & 1 & 6.09 & 0.930 & 0.022 & 2.40  \\ \hline
1.7 & 1 & 12.7 & 0.943 & 0.044 & 4.66  \\ \hline
\end{tabular}
\caption{Redshift bins, fiducial values for $\mu$ and $f$ and their errors for DESI forecast with clustering information, using BGS+ELGs+LRGs+QSOs.}
\label{ta3}
\end{center}
\end{table*}
\begin{table*}[htbp]
\begin{center}
\begin{tabular}{|c|c|c|c|c|c|c|c|c|}
\hline
\multicolumn{9}{|c|}{J-PAS clustering 4000 sq. deg.} \\ \hline
\hline
$z$ & $\mu$ & $\Delta \mu / \mu (\%)$ & $f$ & $\Delta f$ & $\Delta f / f (\%)$ & $f \sigma_8$ & $\Delta f \sigma_8$ & $\Delta f \sigma_8 / f \sigma_8 (\%)$ \\
\hline \hline
  0.30 & 1 & 17.5 & 0.683 & 0.024 & 3.57 & 0.477 & 0.074 & 15.6  \\ \hline
  0.50 & 1 & 7.47 & 0.759 & 0.021 & 2.81 & 0.477 & 0.033 & 6.83  \\ \hline
  0.70 & 1 & 6.14 & 0.816 & 0.023 & 2.84 & 0.465 & 0.027 & 5.75  \\ \hline
  0.90 & 1 & 6.69 & 0.858 & 0.029 & 3.39 & 0.446 & 0.028 & 6.33  \\ \hline
  1.10 & 1 & 8.03 & 0.889 & 0.035 & 3.96 & 0.423 & 0.030 & 7.10  \\ \hline
  1.30 & 1 & 16.9 & 0.913 & 0.068 & 7.42 & 0.400 & 0.052 & 13.1  \\ \hline
  1.50 & 1 & 28.7 & 0.930 & 0.113 & 12.1 & 0.377 & 0.080 & 21.1  \\ \hline
  1.70 & 1 & 30.0 & 0.943 & 0.122 & 12.9 & 0.357 & 0.079 & 22.1  \\ \hline
  1.90 & 1 & 31.9 & 0.954 & 0.132 & 13.9 & 0.337 & 0.079 & 23.5  \\ \hline
  2.10 & 1 & 32.8 & 0.961 & 0.139 & 14.4 & 0.318 & 0.077 & 24.2  \\ \hline
  2.30 & 1 & 39.4 & 0.968 & 0.169 & 17.4 & 0.302 & 0.088 & 29.0  \\ \hline
  2.50 & 1 & 40.8 & 0.973 & 0.177 & 18.2 & 0.287 & 0.086 & 30.0  \\ \hline
  2.70 & 1 & 44.7 & 0.977 & 0.195 & 20.0 & 0.273 & 0.090 & 33.0  \\ \hline
  2.90 & 1 & 49.6 & 0.980 & 0.218 & 22.2 & 0.259 & 0.094 & 36.5  \\ \hline
  3.10 & 1 & 54.9 & 0.983 & 0.242 & 24.7 & 0.248 & 0.100 & 40.4  \\ \hline
  3.30 & 1 & 60.5 & 0.985 & 0.268 & 27.2 & 0.237 & 0.105 & 44.4  \\ \hline
  3.50 & 1 & 67.1 & 0.987 & 0.298 & 30.2 & 0.228 & 0.112 & 49.2  \\ \hline
  3.70 & 1 & 82.2 & 0.989 & 0.363 & 36.7 & 0.218 & 0.130 & 59.7  \\ \hline
  3.90 & 1 & 100  & 0.990 & 0.442 & 44.6 & 0.210 & 0.152 & 72.5  \\ \hline
\end{tabular}
\caption{Redshift bins, fiducial values for $\mu$ and $f$ and their errors for J-PAS forecast with clustering information, 4000 square degrees and using ELGs+LRGs+QSOs.}
\label{ta1}
\end{center}
\end{table*}
\begin{table*}[htbp]
\begin{center}
\begin{tabular}{|c|c|c|c|c|c|c|c|c|}
\hline
\multicolumn{9}{|c|}{J-PAS clustering 8500 sq. deg.} \\ \hline
\hline
$z$ & $\mu$ & $\Delta \mu / \mu (\%)$ & $f$ & $\Delta f$ & $\Delta f / f (\%)$ & $f \sigma_8$ & $\Delta f \sigma_8$ & $\Delta f \sigma_8 / f \sigma_8 (\%)$ \\
\hline \hline
  0.30 & 1 & 12.0 & 0.683 & 0.017 & 2.45 & 0.477 & 0.051 & 10.7  \\ \hline
  0.50 & 1 & 5.12 & 0.759 & 0.015 & 1.93 & 0.477 & 0.022 & 4.68  \\ \hline
  0.70 & 1 & 4.21 & 0.816 & 0.016 & 1.95 & 0.465 & 0.018 & 3.95  \\ \hline
  0.90 & 1 & 4.59 & 0.858 & 0.020 & 2.32 & 0.446 & 0.019 & 4.34  \\ \hline
  1.10 & 1 & 5.51 & 0.889 & 0.024 & 2.72 & 0.423 & 0.021 & 4.87  \\ \hline
  1.30 & 1 & 11.6 & 0.913 & 0.046 & 5.09 & 0.400 & 0.036 & 8.97  \\ \hline
  1.50 & 1 & 19.7 & 0.930 & 0.077 & 8.32 & 0.377 & 0.055 & 14.5  \\ \hline
  1.70 & 1 & 20.6 & 0.943 & 0.083 & 8.84 & 0.357 & 0.054 & 15.1  \\ \hline
  1.90 & 1 & 21.9 & 0.954 & 0.091 & 9.52 & 0.337 & 0.054 & 16.1  \\ \hline
  2.10 & 1 & 22.5 & 0.961 & 0.095 & 9.90 & 0.318 & 0.053 & 16.6  \\ \hline
  2.30 & 1 & 27.0 & 0.968 & 0.116 & 12.0 & 0.302 & 0.060 & 19.9  \\ \hline
  2.50 & 1 & 28.0 & 0.973 & 0.121 & 12.5 & 0.287 & 0.059 & 20.6  \\ \hline
  2.70 & 1 & 30.7 & 0.977 & 0.134 & 13.7 & 0.273 & 0.062 & 22.6  \\ \hline
  2.90 & 1 & 34.0 & 0.980 & 0.149 & 15.2 & 0.259 & 0.065 & 25.0  \\ \hline
  3.10 & 1 & 37.7 & 0.983 & 0.166 & 16.9 & 0.248 & 0.068 & 27.7  \\ \hline
  3.30 & 1 & 41.5 & 0.985 & 0.184 & 18.6 & 0.237 & 0.072 & 30.4  \\ \hline
  3.50 & 1 & 46.1 & 0.987 & 0.204 & 20.7 & 0.228 & 0.077 & 33.7  \\ \hline
  3.70 & 1 & 56.4 & 0.989 & 0.249 & 25.2 & 0.218 & 0.089 & 41.0  \\ \hline
  3.90 & 1 & 68.9 & 0.990 & 0.303 & 30.6 & 0.210 & 0.104 & 49.8  \\ \hline
\end{tabular}
\caption{Redshift bins, fiducial values for $\mu$ and $f$ and their errors for J-PAS forecast with clustering information, 8500 square degrees and using ELGs+LRGs+QSOs.}
\label{ta2}
\end{center}
\end{table*}
\begin{table*}[htbp]
\begin{center}
\begin{tabular}{|c|c|c|c|c|c|}
\hline
\multicolumn{6}{|c|}{Euclid clustering} \\ \hline
\hline
$z$ & $\mu$ & $\Delta \mu / \mu (\%)$ & $f$ & $\Delta f$ & $\Delta f / f  (\%)$ \\
\hline \hline
0.6 & 1 & 4.88 & 0.789 & 0.017 & 2.12  \\ \hline
0.8 & 1 & 3.42 & 0.838 & 0.014 & 1.65  \\ \hline 
1.0 & 1 & 2.64 & 0.875 & 0.012 & 1.32  \\ \hline
1.2 & 1 & 2.60 & 0.902 & 0.012 & 1.31  \\ \hline
1.4 & 1 & 2.46 & 0.922 & 0.011 & 1.19  \\ \hline
1.6 & 1 & 2.67 & 0.937 & 0.012 & 1.23  \\ \hline
1.8 & 1 & 3.58 & 0.949 & 0.014 & 1.50  \\ \hline
\end{tabular}
\caption{Redshift bins, fiducial values for $\mu$ and $f$ and their errors for Euclid forecast with clustering information, using ELGs.}
\label{ta4}
\end{center}
\end{table*}
\begin{table*}[htbp]
\begin{center}
\begin{tabular}{cc|c|c|c|c|}
\cline{3-6}
                       &  & \multicolumn{4}{c|}{$\Delta \mu / \mu (\%)$} \\ \hline \hline
\multicolumn{1}{|c|}{$k$}    & $\mu$ &  Euclid  & DESI & JPAS $8500 \,\, \mbox{sq. deg.}$     & JPAS $4000 \,\, \mbox{sq. deg.}$     \\ \hline
\multicolumn{1}{|c|}{0.024}  & 1 &  7.02 & 8.48 & 8.47 & 12.4  \\ \hline
\multicolumn{1}{|c|}{0.058}  & 1 &  3.49 & 4.59 & 5.09 & 7.41  \\ \hline
\multicolumn{1}{|c|}{0.093}  & 1 &  2.69 & 3.83 & 4.68 & 6.82  \\ \hline
\multicolumn{1}{|c|}{0.127}  & 1 &  2.50 & 3.80 & 5.10 & 7.44  \\ \hline
\multicolumn{1}{|c|}{0.161}  & 1 &  2.69 & 4.37 & 6.43 & 9.38  \\ \hline
\multicolumn{1}{|c|}{0.196}  & 1 &  3.12 & 5.37 & 8.92 & 13.0  \\ \hline
\multicolumn{1}{|c|}{0.230}  & 1 &  3.99 & 7.39 & 15.0 & 21.8  \\ \hline
\multicolumn{1}{|c|}{0.264}  & 1 &  5.34 & 10.7 & 29.6 & 43.2  \\ \hline
\multicolumn{1}{|c|}{0.299}  & 1 &  7.78 & 17.6 & 67.6 & 98.6  \\ \hline
\multicolumn{1}{|c|}{0.333}  & 1 &  1.21 & 32.6 & 153  & 223   \\ \hline
\end{tabular}
\caption{Centers of bins $k_a$ in units of $\mathrm{h / Mpc}$, fiducial values for $\mu$ and their relative errors for Euclid forecast using ELGs, DESI forecast using BGS+ELGs+LRGs+QSOs and J-PAS forecast using ELGs+LRGs+QSOs with 8500 and 4000 square degrees. All for clustering information.}
\label{ta1k}
\end{center}
\end{table*}

\begin{table}[htbp]
\centering
\begin{tabular}{|c|c|c|c|}
\hline
\multicolumn{4}{|c|}{$n_\theta$ values for J-PAS } \\ \hline
\hline
$\delta z$ & LRG & ELG & LRG+ELG  \\
\hline \hline
 0.003 & 0.52 & 2.48 & 3.00   \\ \hline
 0.01 & 2.02 & 6.21 & 8.23   \\ \hline 
 0.03 & 3.25 & 9.07 & 12.32   \\ \hline
\end{tabular}
\caption{$n_{\theta}$ values for J-PAS with different galaxies and redshift errors, in galaxies per square arc minute.}
\label{ta5}
\end{table}

\begin{table*}[htbp]
\centering
\begin{tabular}{ccc|c|c|}
\cline{4-5}
 &  &  & \multicolumn{2}{|c|}{J-PAS lensing} \\ \cline{4-5}
 &  &  & $8500 \,\, \mbox{sq. deg.}$ & $4000 \,\, \mbox{sq. deg.}$ \\ \hline
\multicolumn{1}{|c|}{$z$} & \multicolumn{1}{c|}{$\ell_{\text{max}}$} & $\eta$ & $\Delta \eta / \eta (\%)$ & $\Delta \eta / \eta (\%)$  \\
\hline \hline
 \multicolumn{1}{|c|}{0.1} & \multicolumn{1}{c|}{40}   & 1 & 12.4 & 18.1   \\ \hline
 \multicolumn{1}{|c|}{0.3} & \multicolumn{1}{c|}{130}  & 1 & 7.98 & 11.6   \\ \hline
 \multicolumn{1}{|c|}{0.5} & \multicolumn{1}{c|}{238}  & 1 & 10.6 & 15.4   \\ \hline 
 \multicolumn{1}{|c|}{0.7} & \multicolumn{1}{c|}{366}  & 1 & 23.6 & 34.4   \\ \hline
 \multicolumn{1}{|c|}{0.9} & \multicolumn{1}{c|}{514}  & 1 & 106  & 154    \\ \hline
 \multicolumn{1}{|c|}{1.1} & \multicolumn{1}{c|}{686}  & 1 & -    & -      \\ \hline 
 \multicolumn{1}{|c|}{1.3} & \multicolumn{1}{c|}{884}  & 1 & -    & -      \\ \hline
\end{tabular}\hspace{1in}\begin{tabular}{|c|c|c|c|}

\hline
\multicolumn{4}{|c|}{Euclid lensing} \\ \hline
\hline
$z$ & $\ell_{\text{max}}$ & $\eta$ & $\Delta \eta / \eta (\%)$  \\
\hline \hline
 0.2 & 83   & 1 & 4.21   \\ \hline
 0.4 & 182  & 1 & 4.48   \\ \hline
 0.6 & 300  & 1 & 3.97   \\ \hline
 0.8 & 437  & 1 & 4.72   \\ \hline 
 1.0 & 597  & 1 & 8.10   \\ \hline
 1.2 & 782  & 1 & 20.9   \\ \hline
 1.4 & 994  & 1 & 78.3   \\ \hline
 1.6 & 1240 & 1 & 490    \\ \hline 
 1.8 & 1510 & 1 & -      \\ \hline
\end{tabular}
\caption{Redshift bins, $\ell_{\text{max}}$ values, fiducial values for $\eta$ and relative errors. In left table, errors for J-PAS, using LRG+ELG galaxies with $\delta z=0.03$. We show only errors using ELG+LRG and lensing information. In right table, errors for Euclid using lensing information.}
\label{ta6}
\end{table*}
\begin{table*}[htbp]
\centering
\begin{tabular}{|c|c|c|c|c|}
\hline
 \multicolumn{2}{|c|}{}  & Euclid & $8500 \,\, \mbox{sq. deg.}$ & $4000 \,\, \mbox{sq. deg.}$ \\ \hline
\hline
$\ell$ & $\eta$ & $\Delta \eta / \eta (\%)$ & $\Delta \eta / \eta (\%)$ & $\Delta \eta / \eta (\%)$  \\
\hline \hline
100   & 1 &  5.35 & 10.3 & 15.0     \\ \hline
250   & 1 &  7.78 & 16.7 & 24.4     \\ \hline
400   & 1 &  8.55 & 63.3 & 92.3     \\ \hline
550   & 1 &  15.2 & 360  & 524      \\ \hline
700   & 1 &  42.1 & -    & -        \\ \hline
850   & 1 &  130  & -    & -        \\ \hline
1000  & 1 &  176  & -    & -        \\ \hline
\end{tabular}
\caption{Centers of bins $\ell_a$, fiducial values for $\eta$ and relative errors for J-PAS, using LRG+ELG galaxies with $\delta z=0.03$ and for Euclid using lensing information.}
\label{ta7}
\end{table*}
\begin{table*}[htbp]
\centering
\begin{tabular}{|c|c|c|c|c|c|c|}
\hline
\multicolumn{7}{|c|}{J-PAS clustering + lensing} \\ \hline
\hline
$z$ & $\Delta \eta / \eta_{\,8500} (\%)$ & $\Delta \eta / \eta_{\,4000} (\%)$ & $\Delta \mu / \mu_{\,8500} (\%)$ & $\Delta \mu / \mu_{\,4000} (\%)$ & $\Delta E / E_{\,8500} (\%)$ & $\Delta E / E_{\,4000} (\%)$  \\ \hline
 0.3 & 4.28 & 6.25 & 11.1 & 16.1 & 7.12 & 10.4 \\ \hline
 0.5 & 6.86 & 10.0 & 4.71 & 6.86 & 3.22 & 4.70 \\ \hline
 0.7 & 17.1 & 24.9 & 4.03 & 5.87 & 2.88 & 4.20 \\ \hline
 0.9 & 88.8 & 129  & 4.49 & 6.55 & 3.34 & 4.87 \\ \hline
 1.1 & -    & -    & 5.47 & 7.97 & 3.98 & 5.80 \\ \hline
 1.3 & -    & -    & 11.6 & 16.9 & 7.88 & 11.5 \\ \hline
\end{tabular}
\caption{Redshift bins, relative errors for $\eta$, $\mu$, and $E(z)$ for  J-PAS considering clustering and lensing information (with $\delta z=3\%$ and ELGs+LRGs+QSOs).}
\label{ta8}
\end{table*}

\begin{table*}[htbp]
\centering
\begin{tabular}{|c|c|c|c|c|c|}
\hline
\multicolumn{4}{|c|}{Euclid clustering + lensing} \\ \hline
\hline
$z$ & $\Delta \eta / \eta (\%)$ & $\Delta \mu / \mu (\%)$ & $\Delta E / E (\%)$   \\
\hline \hline
 0.6 & 2.58 & 4.68 & 3.42 \\ \hline
 0.8 & 3.63 & 2.83 & 1.84 \\ \hline 
 1.0 & 6.78 & 2.31 & 1.54 \\ \hline
 1.2 & 17.6 & 2.36 & 1.59 \\ \hline
 1.4 & 66.9 & 2.35 & 1.61 \\ \hline
 1.6 & 415  & 2.60 & 1.74 \\ \hline
 1.8 & -    & 3.54 & 2.27 \\ \hline
\end{tabular}
\caption{Redshift bins, relative errors for $\eta$ and $\mu$ for Euclid, considering clustering and lensing information.}
\label{ta9}
\end{table*}
%\begin{table*}[htbp]
\begin{table}
\centering
\begin{tabular}{|c|c|c|c|c|}
\hline
Survey & $\Delta \mu / \mu (\%)$ & $\Delta \eta / \eta (\%)$ & $\Delta \mu_0 / \mu_0 (\%)$ & $\Delta \eta_0 / \eta_0 (\%)$   \\
\hline \hline
Euclid       & 0.98 & 1.37 & 7.13 & 3.38 \\ \hline
J-PAS $8500$ & 2.08 & 2.89 & 9.66 & 4.58 \\ \hline
J-PAS $4000$ & 3.03 & 4.21 & 14.1 & 6.68 \\ \hline
\end{tabular}
\caption{Relative errors for constant $\mu$ and $\eta$, and $\mu_0$ and $\eta_0$ for Euclid and JPAS (with 8500 and 4000 square degrees), considering clustering and lensing information.}
\label{ta10}
\end{table}
\end{widetext}

\end{document}